\documentclass[useAMS]{mn2e}

\usepackage{ulem} 
\usepackage{graphicx}
\usepackage{amsmath}
\usepackage{rotating}

\title[]{The M\,4 Core Project with \textit{HST\,} -- V.  
         Characterizing the PSFs of WFC3/UVIS by Focus\thanks{
    Based on observations with the NASA/ESA Hubble Space Telescope,
    obtained at the Space Telescope Science Institute, which is
    operated by AURA, Inc., under NASA contract NAS 5-26555, under
    Large Program GO-12911.
}}
\author[J. Anderson \& L.\,R. Bedin] 
{J.\,Anderson$^{1}$ and L.\,R.\,Bedin$^{2}$\\ 
$^{1}$Space Telescope Science Institute, 
                 3700 San Martin Dr., 
                 Baltimore, MD 21218, USA\\
$^{2}$Istituto Nazionale di Astrofisica - Osservatorio Astronomico di Padova,
                 Vicolo dell'Osservatorio 5, Padova, IT-35122\\
   } 
\begin{document} 
\date{Accepted 2017 May 19. Received 2017 May 19; in original form 2017 April 19} 
 
\pagerange{\pageref{firstpage}--\pageref{lastpage}} \pubyear{2017} 
 
\maketitle 
\label{firstpage} 

%
%
% ...
%

\begin{abstract}  
As part of the astrometric \textit{Hubble Space Telescope (HST)} large
program GO-12911,  we conduct  an in-depth  study to  characterize the
point spread  function (PSF) of the  \textit{Uv-VISual channel (UVIS)}
of the  \textit{Wide Field Camera\,3  (WFC3)}, as a necessary  step to
achieve the astrometric goals of the program.  We extracted a PSF from
each of the 589 deep exposures taken through the F467M filter over the
course of a year and find that the vast majority of the PSFs lie along
a one-dimensional locus  that stretches continuously from  one side of
focus,  through  optimal  focus,  to  the other  side  of  focus.   We
constructed a focus-diverse  set of PSFs and find that  with only five
medium-bright stars  in an  exposure it  is possible  to pin  down the
focus level  of that exposure.   We show that the  focus-optimized PSF
does  a  considerably  better  job  fitting  stars  than  the  average
``library'' PSF, especially when the PSF  is out of focus.  The fluxes
and  positions are  significantly  improved over  the ``library''  PSF
treatment.  These results  are beneficial for a much  broader range of
scientific  applications than  simply  the program  at  hand, but  the
immediate use of  these PSFs will enable us to  search for astrometric
wobble in the  bright stars in the core of  the globular cluster M\,4,
which  would indicate  a dark,  high-mass companion,  such as  a white
dwarf, neutron star, or black hole.
\end{abstract} 

 

\section{Introduction}
\label{sec01}

One of the Hubble  Space Telescope's (\textit{HST}'s) great advantages
is  that  it  has  an extremely  stable  point-spread  function  (PSF)
relative to ground-based telescopes.  The fact that \textit{HST} is in
orbit high above  the atmosphere means that the PSF  that is delivered
to  the detector  is free  of turbulence-related  variations: the  PSF
varies more  spatially across  the detector  than it  does temporally.
The spatial  variations are due  to a combination of  geometric optics
and detector-related  features (such as variable  charge diffusion due
to changes in chip thickness, see Krist 2003).

For  the  case  of  the   \textit{Uv-VISual  channel  (UVIS)}  of  the
\textit{Wide  Field   Camera\,3  (WFC3)}   ---on  which   the  present
investigation is  focused---a set of 7$\times$8  fiducial PSFs arrayed
across the  detector is necessary  to map these variations.   In other
words,  the  PSF can  change  significantly  over about  500  detector
pixels. (Sabbi \& Bellini 2013).

Even  though  the  \textit{HST}  PSF   is  not  beset  by  atmospheric
variations, the  fraction of  light in the  \textit{HST} PSF  core can
vary with time by $\pm$3\%  due to ``breathing'', the focus variations
that result as \textit{HST} changes orientation relative to the Sun or
goes into  and out of the  Earth's shadow during its  orbit.  This has
been explored in numerous documents (see Dressel 2014 for a summary).

A proper  understanding of the  \textit{HST} PSF is critical  for many
high-precision   \textit{HST}   studies.    With  an   accurate   PSF,
systematically  accurate positions  can be  measured for  well-exposed
stars  to better  than 0.01\,pixel  ($\sim$0.4\,mas) enabling  a large
number of astrometric applications.  Also, an accurate PSF is critical
for   weak-lensing  analyses   and   deconvolution-type  analysis   of
barely-resolved or blended sources.

One  of   the  reasons  that   the  \textit{HST}  PSF  has   not  been
characterized  very well  thus  far stems  from the  fact  that it  is
undersampled.  When a  detector is not at  least Nyquist-sampled, that
means its pixels  are too wide to capture all  the spatial information
in the  scene that the  telescope is  delivering to the  detector (see
Lauer 1999).  Each  exposure gives us a limited  amount of information
about the  scene, and we  must combine multiple dithered  exposures in
order  to  fully  represent  all  the  information  delivered  by  the
telescope to the detector.  This is equally true for the scene and for
the  PSF:  we  need  to  combine multiple  dithered  images  to  fully
understand either of them.

Unfortunately,  it is  particularly  complicated  to combine  multiple
dithered exposures with  \textit{HST} on account of  its large optical
distortion.   Even dithers  of 10  pixels cannot  be combined  without
careful  attention to  the forward  and reverse  distortion solutions,
since  a shift  of  10.0 pixels  at  the center  of  the detector  can
correspond to 10.2 pixels at the edge.

Anderson \& King (2000, AK00) developed a detailed procedure whereby a
set of  dithered exposures can be  used to extract a  properly sampled
PSF  model  from  a  series  of  undersampled  images.   AK00  further
demonstrate that this PSF can be used to extract accurate and unbiased
positions for stars in a  single exposure.  This initial procedure was
constructed  for \textit{Wide  Field  Camera\,2 (WFPC2)},  but it  has
since  been  generalized   to  two  out  of  three   channels  of  the
\textit{Advanced Camera for Surveys}  (ACS)'s, namely the \textit{High
  Resolution Channel} (HRC) and  the \textit{Wide Field Channel} (WFC,
see Anderson \& King 2004, Anderson \& King 2006) and to both the WFC3
channels  (UVIS   and  the  \textit{Near  Infra-Red,   NIR},  Anderson
2016).\footnote{  All geometric  distortions,  library  PSFs, and  the
  codes that  use these to  extract positions and fluxes  are publicly
  available at http://\-www.stsci.edu/\-$\sim$jayander/ }

The AK00 procedure shows how to  construct a PSF for a particular data
set, and the PSF should in  principle be valid only for the particular
\textit{average}  state  of  the  telescope  during  those  exposures.
Fortunately,  in practice,  it turns  out that  breathing affects  the
\textit{HST} PSF in specific ways that often do not affect our ability
to do  astrometry or photometry (if  we allow for a  possible shift in
zeropoint   of  $\pm$0.03   magnitude   and  \textit{general}   linear
astrometric   transformations).   For   this   reason,  Anderson   has
constructed  a  set  of  ``library''  PSFs  for  various  filters  and
detectors from data-sets  that are well suited  for PSF reconstruction
(a large  number of  high S/N unsaturated  stars and  several dithered
exposures).  These PSFs can be  used to extract differential positions
good to  better than  0.01 pixel and  differential photometry  good to
better than 0.01 magnitude (see Bedin et al.\ 2013).

As such,  many projects do  not require  tailor-made PSFs, but  can be
done with a static archive  of library PSFs.  Other projects, however,
do require more than just simple differential astrometry or photometry
on well-separated stars.   Such more complicated projects  rely on the
PSF either  to do simultaneous  fitting to multiple  overlapping stars
(such as  in crowded  fields or  resolved-binary studies)  or to  do a
deconvolution-type   fitting  of   resolved   objects   (such  as   in
weak-lensing studies  of field galaxies).  For  these purposes, better
PSF models are necessary.

Unfortunately, many of the projects  for which static library PSFs are
not  adequate do  not have  enough bright,  well-exposed stars  in the
field to allow construction of a  full spatially variable model of the
PSF.  (It typically requires over 560 high-signal-to-noise stars in an
image,  i.e.,  enough to  get  10  stars  in  each of  the  7$\times$8
spatially independent PSF zone.)

If the  \textit{HST} PSF varies in  an irregular way from  exposure to
exposure,  then  there will  be  no  way  to  do these  star-poor  yet
PSF-dependent  projects.   However,  if  it  can  be  shown  that  the
\textit{HST} PSFs  fall largely into a  one-parameter family regulated
by  the telescope  focus and  breathing, then  it may  be possible  to
construct a set of library PSFs, parametrized by focus.  In that case,
rather  than needing  enough  stars  in an  image  to  extract a  full
spatially  variable  PSF, it  may  be  possible  to specify  only  one
parameter:  the  telescope  focus.  This  would  allow  high-precision
PSF-based analyses  for a  great many  exposures that  heretofore have
been amenable only to less rigorous analyses.

In this  article, we will determine  whether such a family  exists for
F467M (the main filter of our \textit{HST} large program GO-12911, see
Sect.\,3) and, if so,  how easy it will be to  pinpoint the focus, and
thus specify  the full PSF, for  a given exposure.  If  this procedure
works  for our  data  set in  one  filter, then  perhaps  it could  be
performed for other filters in other data sets as well.

\begin{figure*}
\centering
\includegraphics[width=17cm]{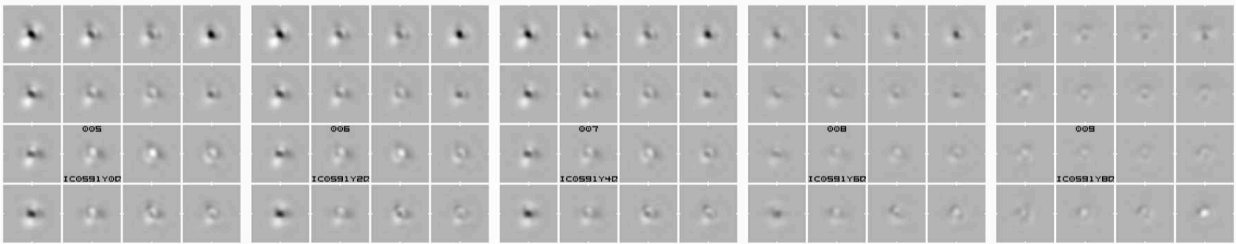}
\caption{The  4$\times$4   perturbation  PSFs  for   five  consecutive
  exposures in the epoch-11 data set.  The exposure name and number in
  the  49-image  series  for  the  epoch are  given  for  each.   Dark
  corresponds to more  flux than the library PSF, white  to less flux.
  The perturbation  PSF changes systematically during  the sequence of
  exposures.}
\label{fig01}
\end{figure*}


\section{The Pilot Study for WFC3/UVIS/F606W}
\label{sec02}

An initial study of how the PSF responds in detail to changes of focus
is documented in Anderson et al.\ (2015, A15).  This study made use of
the fact that the upper-left corner of the detector is known to be out
of focus relative to the rest of the detector, thus making that corner
more sensitive than the rest of the detector to variations of focus.

When the telescope goes  from one side of focus to  the other, the PSF
changes are  symmetric to first  order: some flux is  transferred from
the diffraction rings to the core  and then back again.  To break this
degeneracy, we need to explore  higher-order PSF changes.  Models show
that when the telescope  is on one side of focus,  the PSF is slightly
astigmatic  in  one  sense,  then  on the  other  side  of  focus  its
astygmatism is the opposite.

It is  worth noting that  the undersampled nature of  the \textit{HST}
PSF makes it particularly difficult  to explore this assymmetry, since
all the changes happen within  the star's inner 3$\times$3 pixels, and
the exact location of a star's  center within its central pixels has a
much  larger impact  on  the distribution  of  these 3$\times$3  pixel
values  than does  the  PSF  variation with  focus.   As  such, it  is
critical to model  the average PSF well enough to  examine these small
perturbations  and explore  how the  higher-order aspects  of the  PSF
change with focus.

The investigation in A15 went through the entire WFC3/UVIS archive and
identified all  the exposures that had  a good number of  stars in the
focus-sensitive  upper-left corner.   They isolated  those stars  that
were  centered nearly  perfectly on  their central  pixels so  that it
would be easy to construct metrics to measure the residual astigmatism
using a simple moment-based analysis.
Next, they fit each of these stars with a pre-existing ``library'' PSF
for  the F606W  filter so  that they  could construct  residuals which
could  be  distilled   into  a  relative  sharpness   and  a  relative
astigmatism.
When the authors of A15  plotted the astigmatism against the sharpness
of  the  PSF,   they  found  that  the  PSF  varied   along  a  simple
``banana''-like  path in  this two-parameter  space (see  Figure 7  of
A15).

The fact  that there  is such a  tight empirical  relationship between
astigmatism and sharpness  means that the PSFs apparently  come from a
simple one-parameter family.
It was further found that all of the star images from a given exposure
clustered  in  a single  location  along  this  curve and  stars  from
different exposures clustered in different  locations.  This led us to
conclude that  the different locations  along the curve  correspond to
different focus levels of the telescope.

The next step was to group  together exposures that had the same focus
and extract images of unsaturated  but well-exposed stars (i.e., those
with between  5$\times$10$^4$ and  3$\times$10$^5$ total  counts) that
were in the  upper-left corner.  A15 then used  these extracted images
of actual stars to construct an average PSF for the upper-left corner,
one for  each of  eight different  focus levels,  thus arriving  at an
empirical picture of how the PSF changes.  Figure 13 of A15 shows this
variation  visually.   It  is  clear  that the  PSF  goes  from  being
asymmetric in the $+135^{\circ}$ direction at their first focus level,
to being asymmetric  in the $+45^{\circ}$ direction at  the other side
of focus, at their focus-level 8.

This limited  study of the F606W  PSF in the upper-left-corner  of the
WFC3/UVIS detector  was an  encouraging indication  that it  should be
possible to characterize  the PSF across the entire  detector in terms
of focus.  This is clearly the next step to take.


\section{The next step}
\label{sec03}

Because  of the  undersampled  nature  of the  PSF,  we  need to  have
multiple observations of point sources at multiple sub-pixel locations
if we  hope to construct an  accurate super-sampled model of  the PSF.
The fact that the PSF changes spatially across the detector means that
we need many stars in  each spatially-coherent zone.  Anderson \& King
(2000)  showed that,  in  addition to  this,  we also  need  a way  to
determine an accurate position for each PSF-contributing star.  All of
this means that we require multiple dithered images, each of which has
a large number  of stars.  Globular clusters have served  as the ideal
targets for this purpose.

As such,  large program  GO-12911 (PI:  Bedin) is  perfect for  such a
study.  Its main focus is to  do high-precision astrometry of stars in
globular  cluster M4  over the  course  of 12  months with  an aim  to
measure the  astrometric wobble of  main-sequence stars that  may have
heavy unseen binary companions (such as black holes, neutron stars, or
white  dwarfs).   Bedin  et  al.\ (2013,  Paper\,I)  provides  for  an
overview.

The observations  are divided into  12 epochs, spaced roughly  a month
apart.   Each  epoch  consists  of  $\sim$49  deep  observations  with
WFC3/UVIS through the medium-band F467M filter, wherein the turnoff is
just below the level of  saturation, thus providing the maximum number
of high  signal-to-noise stars.  The  F467M filter was chosen  so that
the  ideal  exposure time  would  be  just  over  339s, which  is  the
threshold for  efficient buffer-dumping  with WFC3/UVIS.   The program
also took short  exposures in filter F775W to provide  colors, as well
as short  exposures in  F467M to  provide some  handle on  the evolved
population.  But we focus here on the 589 deep exposures through F467M
taken  in  120 single-orbit  visits  between  9  October 2012  and  16
September 2013.


\section{Grouping the observations by focus}
\label{sec04}

\subsection{Using all the stars}
\label{ssec04.1}
Even  though the  upper-left corner  is more  sensitive to  changes in
focus than  the rest of  the detector, in  reality the PSF  across the
entire detector  changes when the  focus changes, and there  is surely
more information  in the entire detector  than in just the  upper left
sixteenth  of  the   detector.   It  is  hard,  though,   to  use  the
moment-based analysis that worked in  the upper-left corner across the
entire  chip, since  different locations  across the  detector are  at
different places  along the  local focus curve.   For this  reason, we
sought a way to extract an estimate of the full spatially variable PSF
for each exposure.

We    noted   above    that    we   would    need    at   least    560
well-exposed-but-unsaturated stars in an image if we hope to constrain
the  array of  7$\times$8 PSFs  with a  minimum of  ten stars  in each
fiducial-PSF region.   This is a lot  to ask for, even  in the central
field of a globular cluster, where  the center of the cluster is often
too dense to  hope for isolated stars and the  outskirts too sparse to
get enough  stars.  For this reason,  we decided to first  construct a
time-averaged ``library''  F467M PSF that  included the full  array of
7$\times$8 fiducial PSFs  and use that as the basis  for the PSF model
in each exposure.  Then for  each exposure, we constructed a ``delta''
PSF to  describe how the  PSF in that  exposure is different  from the
time-averaged basis PSF.  Such an approach allows the well-constrained
average model to deal  with the fine-scale detector-related variations
(traceable to issues such as  static distortion, or chip thickness and
charge diffusion or vignetting) while  at the same time accounting for
the low-spatial-frequency variations due to focus.

The ``basis'' PSF  model for WFC3/UVIS is similar  to that constructed
in 2006 for ACS/WFC in Anderson \& King (2006), except that instead of
using an array  of 9$\times$5 PSFs across  each 4096$\times$2048 chip,
we adopt  a 7$\times$4 array  since that better matches  the structure
present in the UVIS PSF.  In particular, this array placement allows a
fiducial PSF to be centered on  the ``happy bunny'' location where the
PSF is  the sharpest (see  Sabbi \& Bellini  2013, SB13).  The  PSF at
each  of the  56  (7$\times$8) fiducial  locations  is represented  by
101$\times$101 super-sampled grid, as in  AK00.  The super sampling is
$\times$4 with  respect to the  image pixels,  so that the  entire PSF
goes out to  about 12.5 pixels from  the center of a  star.  Even with
such dense sub-sampling,  it is necessary to use a  bi-cubic spline to
evaluate the PSF at locations in-between the grid points.

The net empirical PSF that we  construct for each exposure consists of
the average PSF  described above plus a  spatially variable 4$\times$4
perturbation tailored  to better fit  the bright stars in  t exposure.
The perturbation PSF for each exposure is constructed by identifying a
number  (typically a  hundred or  more) of  bright and  isolated stars
across  the image.   Each of  these bright  stars is  fitted with  the
``library'' PSF  appropriate for  its location in  the image,  and the
model PSF  is subtracted,  leaving an array  of scaled  residuals with
respect  to  the  center  of  each star.   These  residuals  show  the
difference  between the  image's true  PSF  and the  library PSF.   We
distill  these residuals  into an  4$\times$4 array  of fiducial  PSFs
(with fiducial PSFs  at the corners and edges, so  that there is never
any extrapolation).  Figure~\ref{fig01} shows the perturbation part of
the PSF for five exposures in the eleventh epoch.

This 4$\times$4 array of perturbations  was combined with the original
library PSF  to construct  a full  7$\times$8 array  of PSFs  for each
exposure.  The aim  is then to see  whether or not this  group of PSFs
covering  the  full  year  of  exposures can  be  generalized  into  a
1-parameter family.

We naturally  start out  not knowing  the focus level  for any  of the
exposures.  Although there  is some engineering data that  may be able
to estimate  the PSF focus  (Cox \& Niemi 2011),  it is not  clear how
accurate the PSF predictions are, and it  is not clear how to tie them
to  accurate  spatially variable  models.   It  is possible  that  the
relative-focus measurements  we extract  from this  paper may  make it
possible to make better use of engineering data.

The task of generalizing these 589 individual-exposure PSF models into
a  family  might lend  itself  to  some  kind of  principal  component
analysis, but it  is not clear that the variations  due to focus would
be linear.  So, instead of exploring that strategy we sought something
simpler and more empirical.


\subsection{``Phylogram'' plots}
\label{ssec04.2}

\begin{figure*}
\centering
\includegraphics[width=17cm]{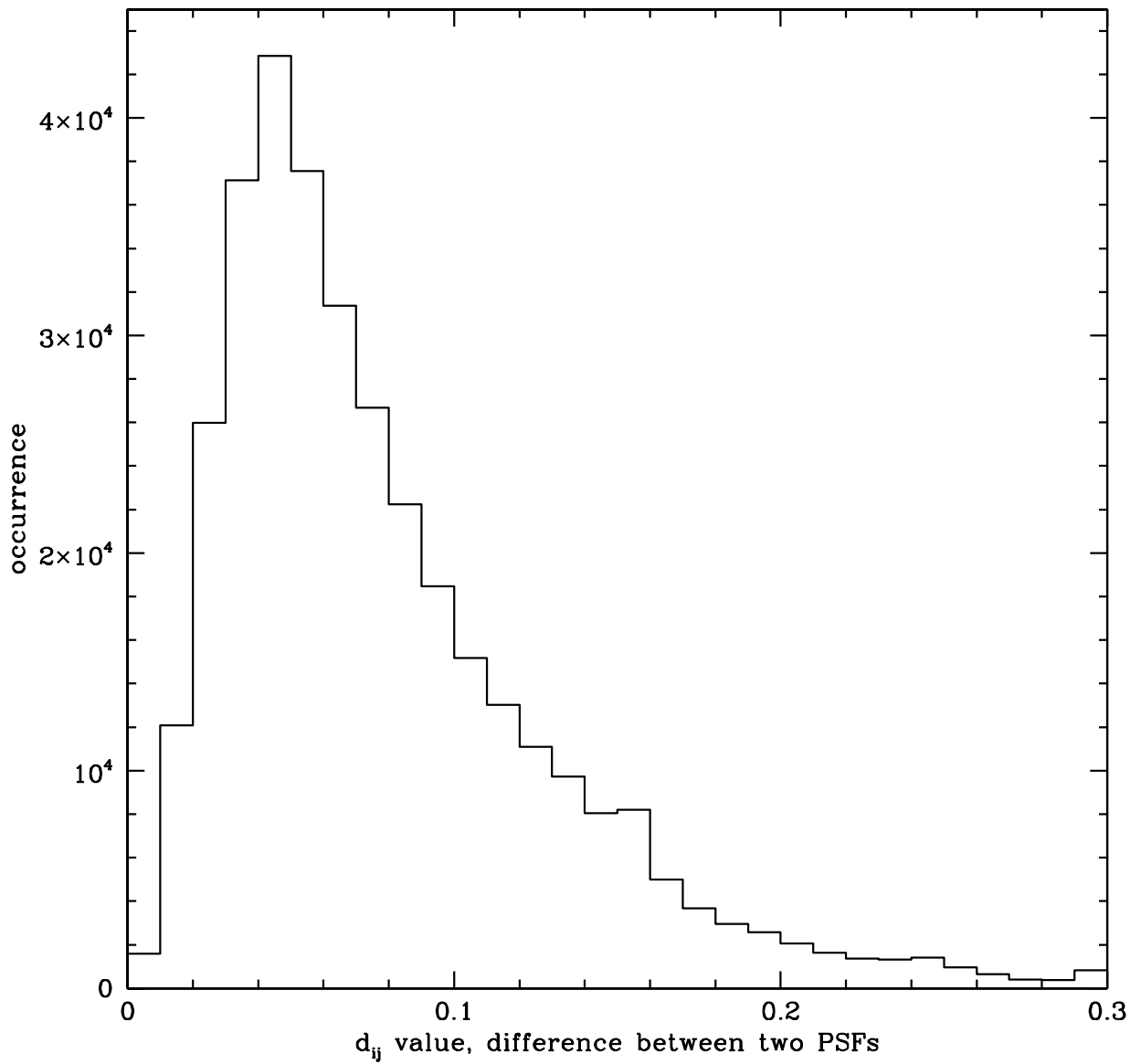}
\caption{The distribution of values for $d_{ij}$, which represents the
  difference between the PSFs  extracted from two different exposures.
  A value of 0.1  means that 10\% of the flux from  one PSF would have
  to be re-distributed to arrive at the other PSF. }
\label{figA}
\end{figure*}

We compared the 7$\times$8 array of PSFs from each individual exposure
against  the  corresponding-zone  PSFs   for  every  other  individual
exposure.   To  do  this,  we  computed  the  absolute  value  of  the
difference between  the 7$\times$8 arrays  of PSFs for images  $i$ and
$j$ as:
$$
         d_{ij} =  \frac{1}{56} \times 
                   \sum_{IJ} \bigl( 
                   \sum_{XY} | \psi_{IJXY;i} - \psi_{IJXY;j}| \bigr) 
                   \ \ \ (1),
$$
where X and Y go from 1 to 101 covering the $\Delta x$ and $\Delta y$ 
domain of the PSF, and $I$ ($J$) goes from 1 to 7 (8), corresponding to 
the spatial variation of the PSF.
These simple difference estimates should  tell us which exposures have
PSFs that  are more similar to  each other (small $d_{ij}$)  and which
have  PSFs  that  are  are  more  different  from  each  other  (large
$d_{ij}$).    Figure\,\ref{figA}  shows   the   distribution  of   the
589$\times$589 ($\sim$350\,000) $d_{ij}$ values.   Most PSFs differ by
about 0.05 (equivalent a  5\% shift of the flux from  one place in the
PSF to another), but there is  a significant tail that differs by more
than 10\%, and some that differ by more than 20\%.

In the field  of biology, it is common to  characterize the difference
between two  species in terms of  the total ``difference'' of  the DNA
code that represents them.  Multiple  species can be inter-compared on
a ``phylogram'',  a two-dimensional  plot that  represents graphically
the multi-dimensional  differences among the various  species in terms
of differences between points in a 2-dimensional space.  Such diagrams
show, for instance,  how similar humans are  to chimpanzees, dolphins,
and paramecia.

In the present  study, we also have  a large number of  data sets that
have  been characterized  by  their differences  in some  quantitative
domain.   It is  worth  investigating whether  such a  two-dimensional
graph could be useful in this case as well.

We  explored  several approaches  to  constructing  such an  optimized
phylogram-type  plot.  Of  course  no two-dimensional  plot  can do  a
perfect job representing the myriad of differences among the PSFs, but
our aim  is to construct  the best possible two-dimensional  plot that
represets these  differences.  In  phylogram-type plots,  the distance
between  species  is  representative  of  the log  of  the  number  of
differences in their DNA sequences.  Here, we seek a diagram where the
two-dimensional  ``distance''   between  two  PSFs  in   our  plot  is
representative of $d_{ij}$, the difference computed above.

\begin{figure*}
\centering
\includegraphics[width=17cm]{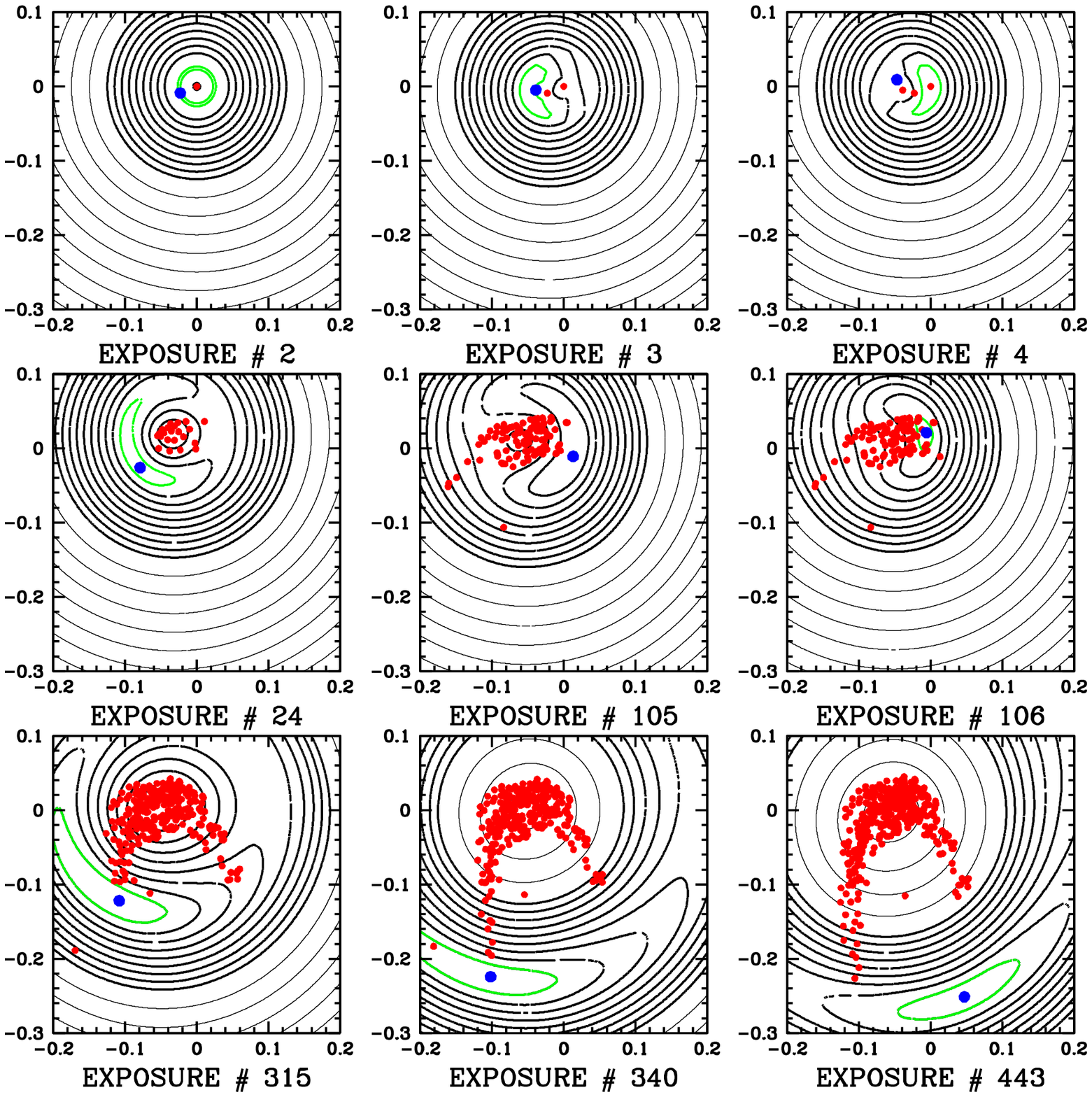}
\caption{The panels show  the construction of the  phylogram by adding
  one  expsure at  a  time.   Each panel  shows  the previously  added
  exposures in red.  The contours  identify the best location to place
  the current  exposure with  respect to  the previous  exposures such
  that the  computed $d_{ij}$  (the distance between  the PSFs)  is as
  close as  possible to $r_{ij}$  (distance between the points  on the
  graph).  The axes are arbitrary,  but their scale represents that of
  $d_{ij}$, namely how much flux would have to be moved to go from one
  PSF to another.  }
\label{figB}
\end{figure*}

We devised several stragies for coming up with such a diagram and show
one such  strategy in  Figure~\ref{figB}.  The goal  is to  place each
point such  that its  distance (in  $\Delta$PSF terms)  represents how
different  the PSFs  are.  The  $x$  and $y$  axes in  this space  are
arbitrary, but  the distance between  two points is  representative of
how similar the PSFs are.

We start  with the representative of  how different the PSFs  are.  We
start with the first exposure at ($x_1$,$y_1$) = ($0.$,$0.$).  For the
second exposure, we explore all possible trial locations ($x_2$,$y_2$)
between  $(-0.3:0.3,-0.3:0.3)$,  spaced  by 0.001  and  determine  the
difference between $d_{12}$  (which happens to be  0.025) and $r_{12}$
(which we define to be $\sqrt{(x_2-x_1)^2+(y_2-y_1)^2)}$).
We identify the minimum of $E_2 \equiv |d_{12}-r_{12}|$ to be the best
location  to place  the second  exposure  relative to  the first;  the
quantity  $E$  represents  the  difference  between  the  distance  as
measured in the plot and the  distance measured between the PSFs.  The
upper left plot in Figure~\ref{figB} shows the contours of $E$.  There
is a circle (shown in  green) of best-placement locations for exposure
\#2, since  $d_{12}$ equals $r_{12}$  along the circumference  of this
circle, and $E_2$ has its minimum value of 0.0.
We  adopt an  arbitrary point  in this  circle as  the best  placement
location for exposure \#2.

Once exposure  \#2 is  placed, we  can perform  the same  operation to
determine where best to place the third exposure.  The middle panel on
the  left   shows  the   contour  plot  for   $E_3$  (defined   to  be
$\sum_{n=1}^{3-1} |d_{n3}-r_{n3}|$).
The previous points  are shown in red and the  lowest contour is shown
in  green.  The  third  exposure  (shown in  blue)  is  placed at  the
location with the lowest value of $E_3$.
After each  new placement,  we explore  the local  neighborhood around
each  point to  see whether  it might  have a  lower value  of $E$  by
shifting by 0.001 in any direction.

\begin{figure*}
\centering
\includegraphics[width=17cm]{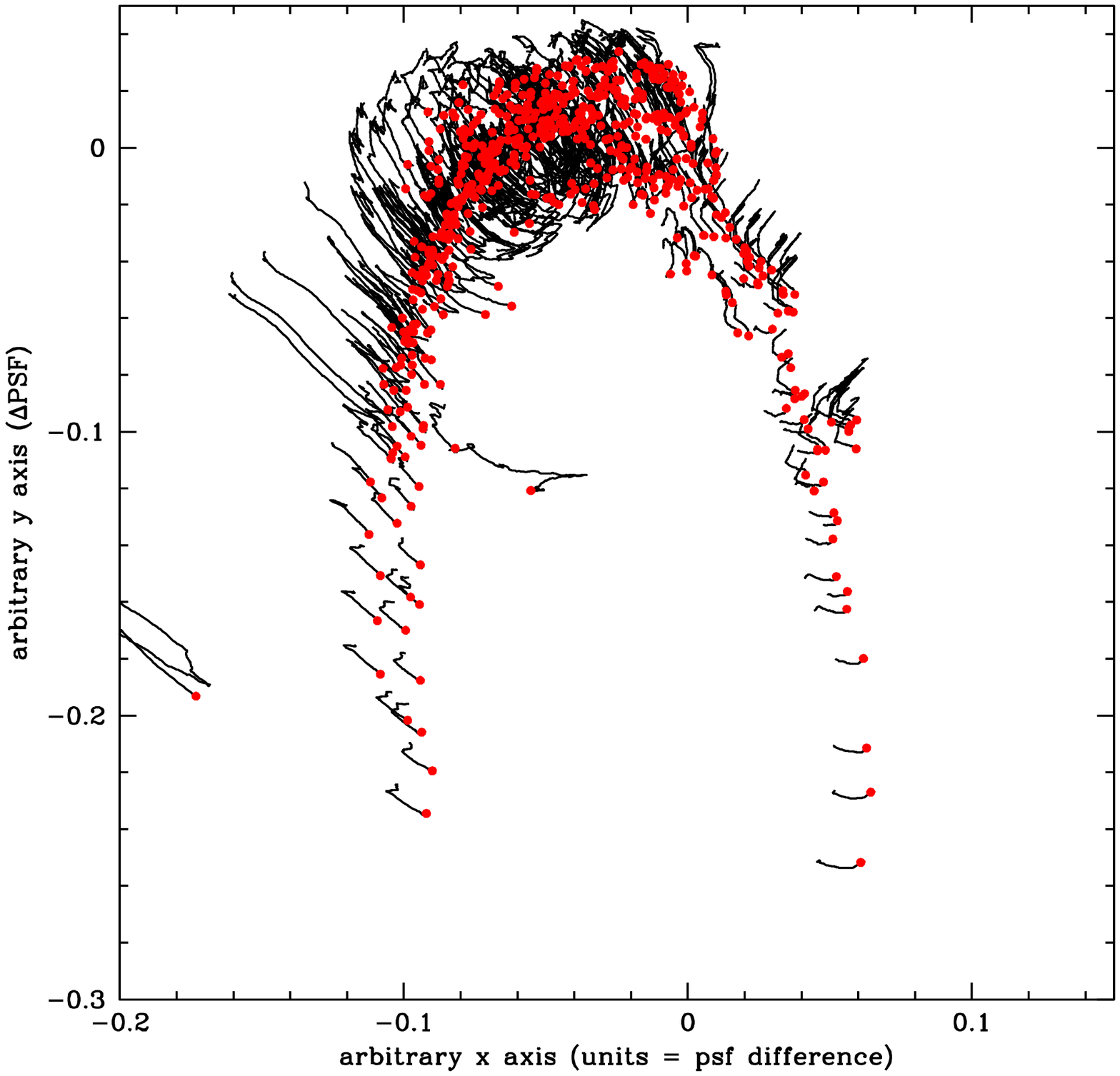}
\caption{The panels  show how  the placement of  the exposures  in the
  phylogram plot change  as more point are added to  the diagram.  The
  open end  of each  black curve  show the  initial placement  of each
  exposure (based on the preceding exposures),  and the red dot at the
  other end shows  the final, optimal resting place  for the exposure,
  after its position has been allowed to creep slowly as more and more
  points are added.}
\label{figC}
\end{figure*}

The other  panels in  Figure~\ref{figB} show  the optimal  location to
place  exposures \#4,  \#24, \#105,  \#106, \#315,  \#340, and  \#443,
based on all the previous placements.   It is clear that the exposures
naturally self-organize into a relatively orderly sequence.
Figure~\ref{figC} shows how the placement  of each exposure evolves as
subsequent exposures are added.
We explored several different starting positions for all the stars and
several different orders for adding the  stars and they all produced a
qualitatively similar diagram (modulo a rotation).

\begin{figure*}
\centering
\includegraphics[width=17cm]{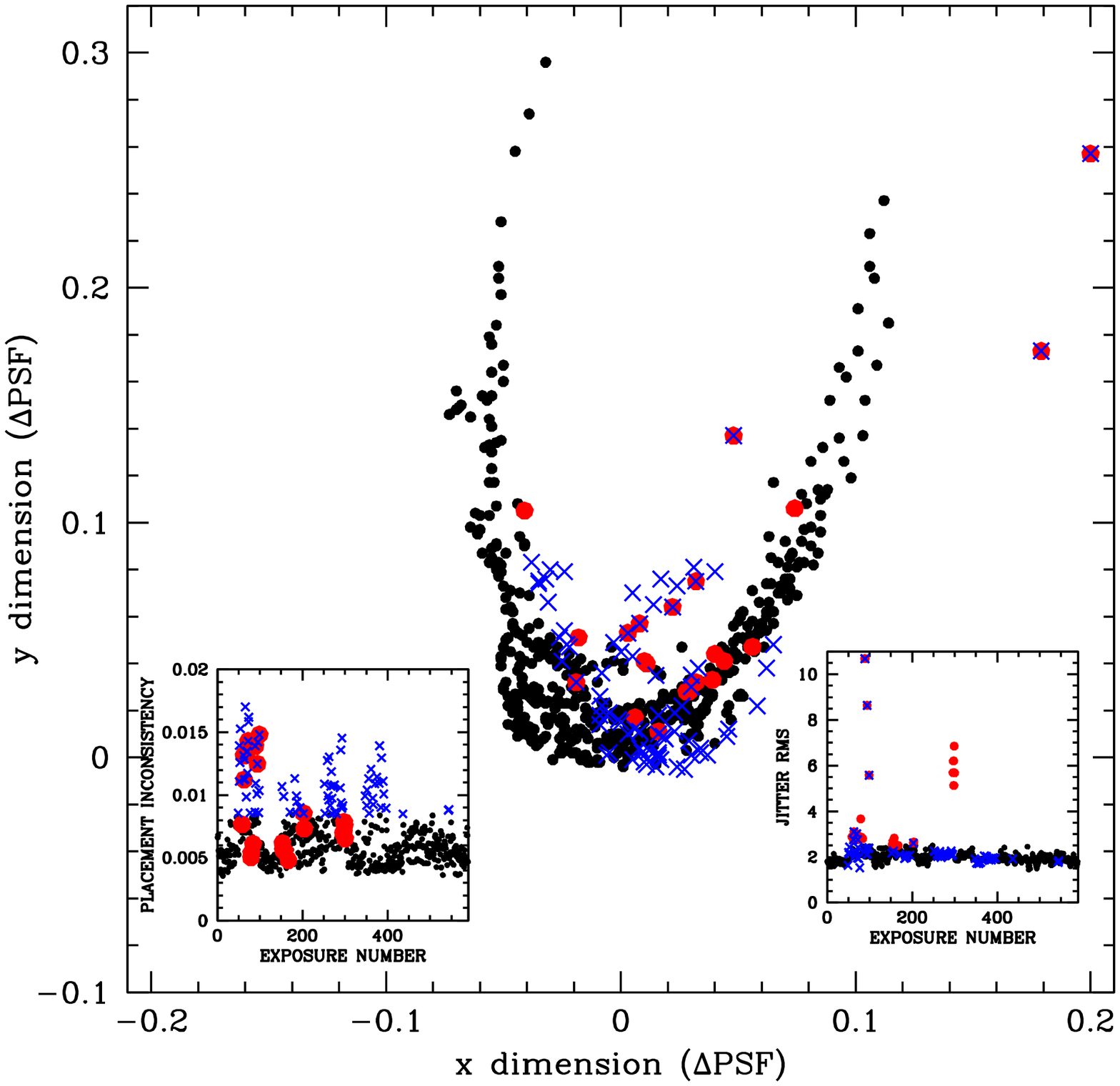}
\caption{This  the  final  reference phylogram-type  plot  where  each
  exposure is  represented by  a dot,  and the  dots are  separated by
  ``distances'' that correspond to  the difference between their PSFs.
  The  axes are  arbitrary, but  units for  this are  in terms  of the
  average absolute  difference between  the PSFs ($d_{ij}$,  such that
  0.1 means that to get from PSF to another you have to rearrange 10\%
  of the  flux.  The inset at  the lower left shows  $E_{\rm min}$ for
  each  exposure.   Stars  with  low  $E_{\rm  min}$  have  consistent
  placement with respect  to most of the other  exposures.  Points are
  marked blue if  they have a high value of  $E_{\rm min}$.  The inset
  in  the  lower  right  shows  the  jitter  RMS  for  each  exposure.
  Exposures with more pointing jitter than typical are flagged in red.
}
\label{fig02}
\end{figure*}

In the end adopted Figure~\ref{fig02}  as our reference phylogram.  We
have re-centered  and oriented  the points  for ease  of presentation.
The 589 points  are situated such that their distance  from each other
best represent  the difference in  the respective PSFs.  The  shape of
this curve  is interesting.   It is  not simply  a straight  line, but
rather it has a  strong curvature to it.  This means  that the PSFs at
the ends are more different from the  PSFs in the middle than they are
from each other.  This is in agreement  with what we saw in the banana
plots from Figure 7 of A15: when the PSF is in focus, it has a maximal
fraction of its flux in its central pixel, but this fraction goes down
in a similar  way on both sides  of focus.  The PSF  is not identical,
however, on  the different  sides of focus,  as indicated  by distance
between the two sides of the wishbone.

Since there are a few outliers  from the trend, we investigate several
possible causes.  We  extracted the jitter file for  each exposure and
determined an RMS with respect to the average pointing.
In Fig.\,\ref{fig02} observations with large  jitter are shown in red.
We also determined which of the observations were able to be placed in
a location on the plot that  was a good represtation of the difference
in their PSFs.  The two-dimensional  plot allowed most observations to
be  placed  such  that  $E  \equiv \sum  |d-r|$  is  small,  but  some
observations (see left inset) were not.
Perhaps this  is an  indication that a  third dimension  might exhibit
even more  order, but since  these observations  are few, we  chose to
ignore them and focus on the those that followed the majority trend.

\begin{figure*}
\centering
\includegraphics[width=17cm]{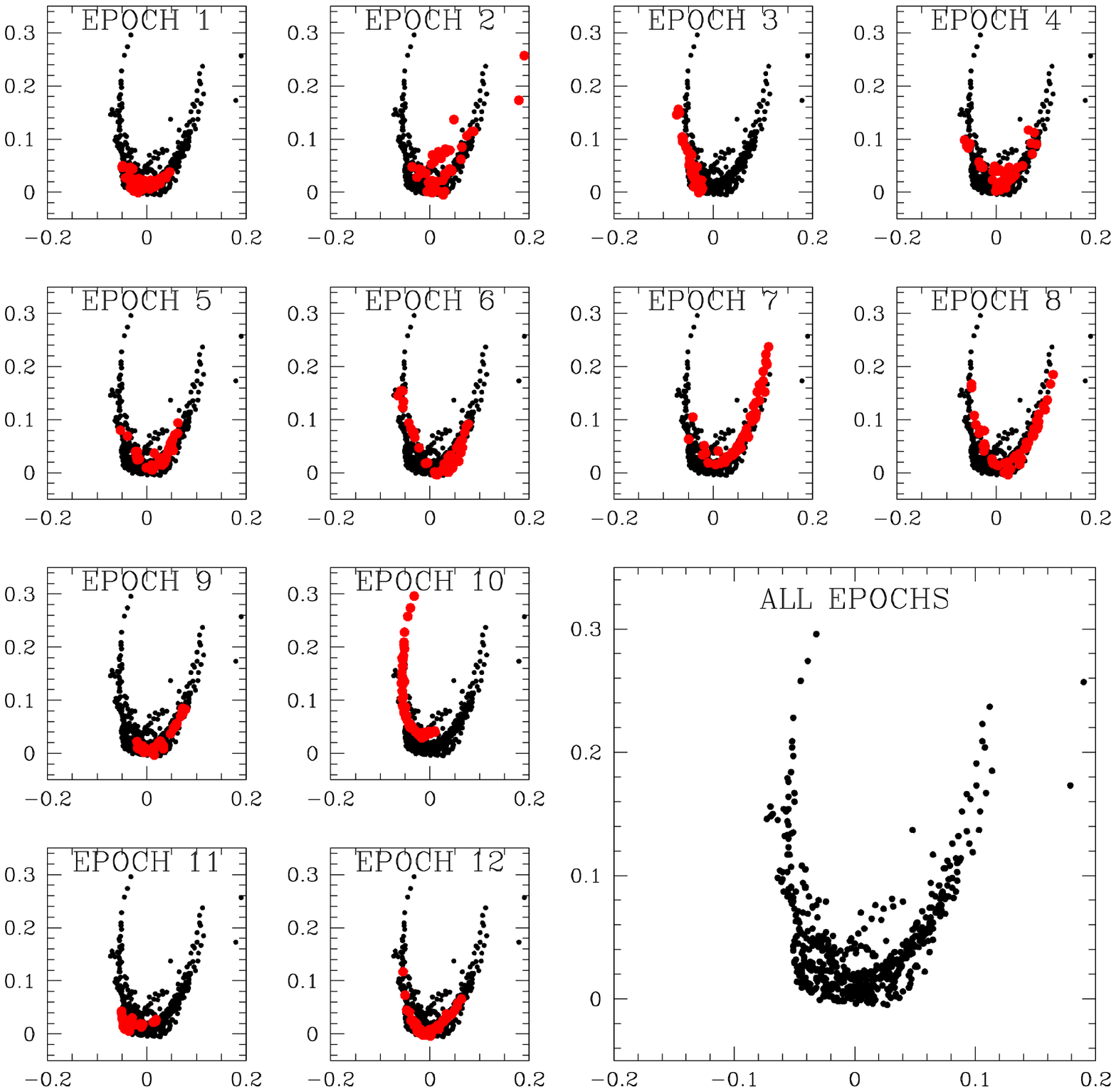}
\caption{This   plot  shows   in   black  the   same   points  as   in
  Figure~\ref{fig02}, but for each of  the 12 epochs it highlights the
  49 exposures taken during that epoch. }
\label{fig03}
\end{figure*}

Figure~\ref{fig03} shows the same phylogram-type  plot, but in each of
the small panels we highlight  in red the observations that correspond
to a particular epoch.  The epochs  are spaced about a month apart and
the visits  within each  epoch span  between 21 and  45 hours.   It is
clear that the PSF  is not constant over an epoch,  though the PSFs do
typically vary over a relatively narrow part of the entire focus range
during an epoch.
There is no clear progression over time from month to month.  

Several visits in  Epoch 2 do not follow the  focus curve.  Inspecting
the  images, we  see that  several  of them  suffered from  guide-star
failures.  Indeed,  this is borne out  by an inspection of  the jitter
files.  All the other epochs follow the well-worn focus path to within
0.02  (corresponding to  a 2\%  average difference  between the  PSFs,
meaning  that  to get  from  one  PSF to  another  one  would have  to
rearrange about 2\% of the flux).

\begin{figure}
\centering
\includegraphics[width=8.4cm]{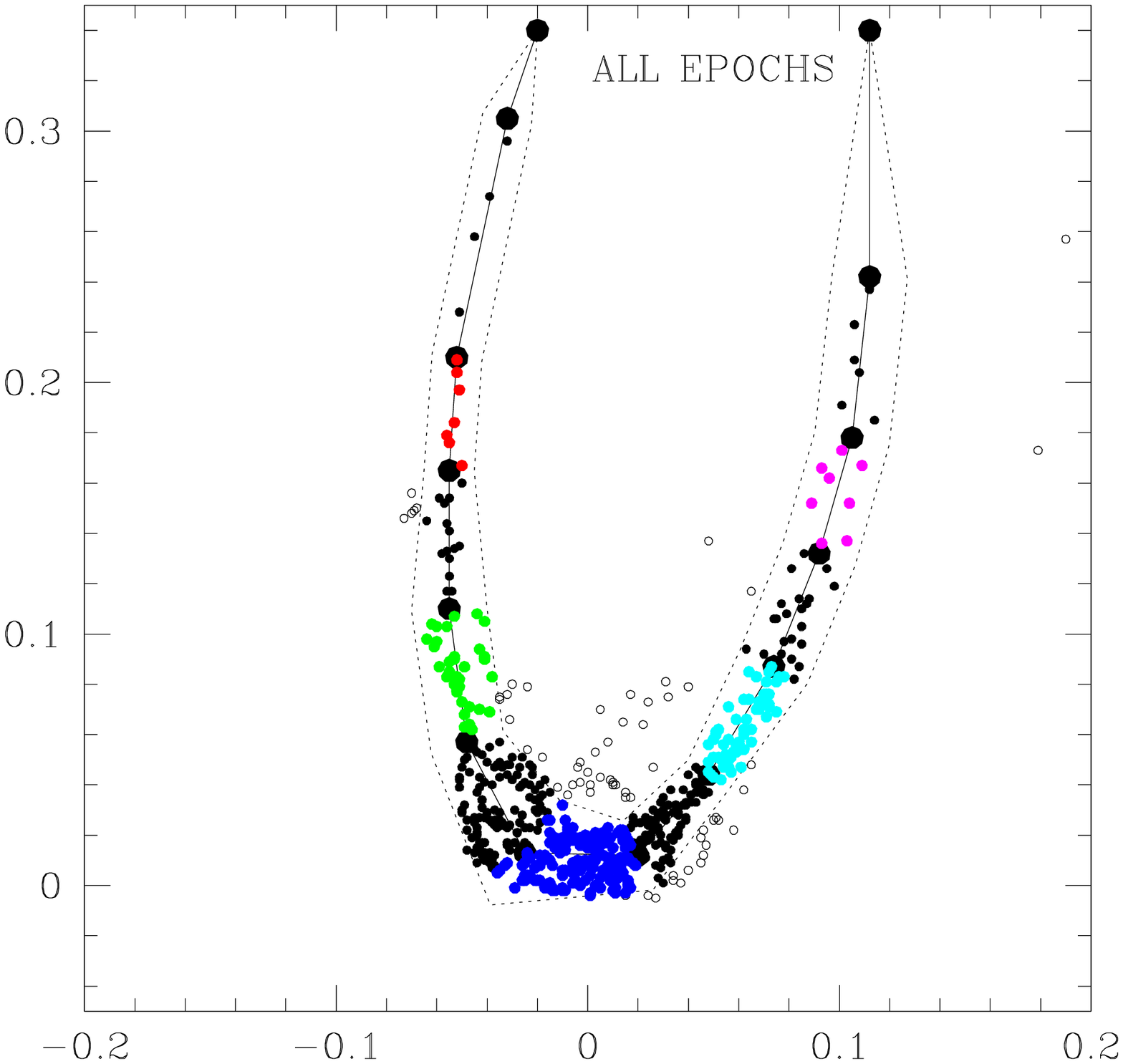}
\caption{This plot shows the  specified zones along the phylogram-type
  focus curve. }
\label{fig04}
\end{figure}

Now that we are able to  characterize individual exposures in terms of
where they lie along the focus  curve, we can group together exposures
at similar focus levels.  Figure~\ref{fig04}  shows the same points as
in  Figure~\ref{fig02},  but  this  time  we  have  color-coded  those
observations at each focus level.
We have drawn in a fiducial line for the focus curves and consider all
observations within about 0.02 (98\% PSF agreement) of the focus curve
to  be  representative.   We  arbitrarily divide  the  curve  into  11
distinct  focus  zones and  color-code  black  the  odd zones  and  in
different  colors the  even ones.   The  open circles  denote the  few
observations that  did not follow the  general focus trend as  well as
the others  (mostly during second  epoch) that suffered  major guiding
failures.
The first focus  group had 4 exposures  and the last group  5, but the
other groups had between 7 and 150 representative exposures.
Focus  groups   at  the  extremes  naturally   have  fewer  exposures,
reflecting the fact that most of the times the telescope is on focus.

\subsection{Finding a PSF for Each Focus Level}
\label{ssec04.3}
The next  step was to take  all the exposures associated  with a given
focus group and  determine an average PSF for each  group.  AK00 shows
that in order to derive an accurate PSF from stellar profiles, we must
have accurate positions and fluxes for each star in each exposure.  In
undersampled  images,  it  is  hard  to  determine  accurate  unbiased
positions, so  AK00 developed a  procedure to  take a dithered  set of
data and  iterate between the solution  for PSF and for  the positions
and fluxes of the stars in a virtuous cycle.

Here, we simply determined a position  for each star using the average
``library'' PSF and  determined the flux by means of  the total amount
of light  within a 5-pixel  radius.  This is considerably  larger than
the 5$\times$5-pixel fitting  region, since we wanted  to include flux
that landed outside of the fitting region.

With   positions   and  fluxes   in   the   unresampled  {\tt   \_flc}
images\footnote{
The   {\tt  \_flc}   images  are   produced  by   the  STScI   archive
pipeline.  They  are  {\tt  \_flt}  images  corrected  from  imperfect
charge-transfer efficiency  (CTE) with an algorithm  essentially based
on the one presented in Anderson \& Bedin (2010).
},
we can  extract a  spatially variable  PSF from  the set  of exposures
associated with  each focus  zone.  As  mentioned earlier,  there were
between 4 and 150 exposures in each zone.

\begin{figure}
\centering
\includegraphics[width=8.9cm]{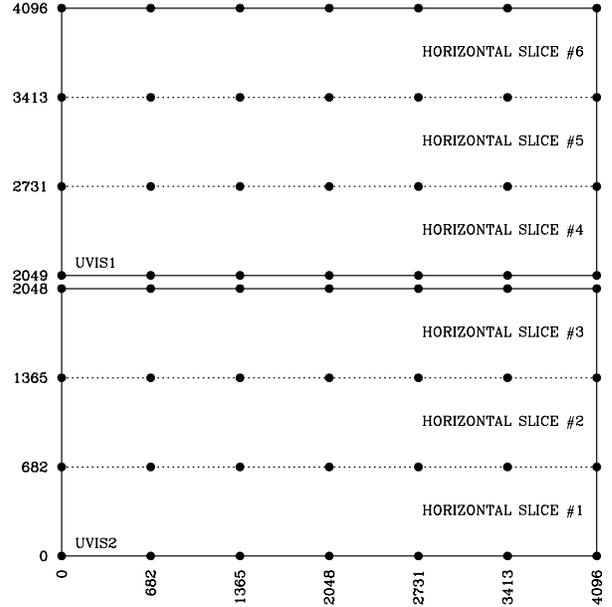}
\caption{ This  plot shows the  locations of the fiducial  PSFs across
  the  two  WFC3/UVIS detectors.  The  useful  part  of each  chip  is
  4096$\times$2048  pixels  and  for  each  of the  two  an  array  of
  7$\times$4 PSFs  is sufficient to  map the spatial  variation; i.e.,
  for a  total of  7$\times$8 PSFs  across the  entire field  of view.
  Note  that there  are  PSFs at  the  edges and  corners  so that  no
  extrapolation  will ever  be necessary.   The PSF  is allowed  to be
  discontinuous across the gap between the chips. The horizontal-slice
  labels pertain to the next plot.}
\label{fig05}
\end{figure}

Since we are  now extracting detailed PSF models, it  is beneficial to
illustrate  specifically where  each fiducial  PSF is  located on  the
detector.  As we mentioned before, each chip is covered by an array of
7$\times$4 fiducial PSFs,  as such the PSFs are spaced  by 682 pixels.
We place PSFs at the edges and  corners of each chip in order to avoid
the  need for  extrapolation.   The PSF  is  linearly interpolated  in
between fiducial locations but is  not assumed to be continuous across
the intra-chip  gap.  Figure~\ref{fig05}  shows the locations  for the
fiducial PSFs for our UVIS PSF model.

The next figure  (Figure~\ref{fig06}) provides a snapshot  of the data
that went into our PSF models.  From  AK00, the PSF model tells us the
fraction of a star's light that  should fall in pixel $[i,j]$ relative
to its  center at $(x_*,y_*)$.   We can thus  model the star  with the
following equation:
$$
         P_{ij} = z_* \times \psi(i-x_*,j-y_*) + s_*,
$$
where $z_*$  and $s_*$ are  the star's  total flux and  background sky
value, respectively.  This is the equation for a line, where the slope
is the flux and the intercept is the sky.  We can invert this to solve
for the  PSF at a single  point in its domain  ($\Delta x$,$\Delta y$)
from the value of pixel $[i,j]$:
         $$
         \psi(\Delta x,\Delta y) = (P_{ij}-s_*)/z_*, 
         $$
where $\Delta x$=$i-x_*$ and $\Delta y$=$j-y_*$.  Each pixel in each
star's image thus provides an estimate of the PSF at one particular 
location in its domain.

When we allow for spatial and focus variations, we see that the PSF is
now a 5-dimensional function: $\psi(\Delta x, \Delta y; i, j; f)$.  To
visualize  it,  we  will  consider  two  dimensions  at  a  time.   In
Figure~\ref{fig06}, we  plot the  samples from the  center of  the PSF
($|\Delta x|<0.25$ and  $|\Delta y|<0.25$) for the  middle focus level
($f$=6) as  a function of  detector $i$ coordinate for  six horizontal
slices across the detector (shown in Fig~\ref{fig05}).

We see  that even  for the  ``optimal'' focus  level, the  fraction of
light in  the central pixel  can vary from  0.175 to 0.225,  more than
$\pm$10\%.  There  is clear structure on  $\sim$500-pixel scales.  The
peak  at the  bottom  of the  detector corresponds  to  a region  with
enhanced
fringing called the ``happy bunny'' (SB13).

The solid  blue dots and  connecting lines  show the actual  PSF model
across each  strip.  The  dark-blue line does  not represent  the data
perfectly, but it is good to better than 0.5\%.

On the right, we show the  same connected points for the central-focus
sample  ($f$=6  in dark  blue,  as  on the  left)  and  also show  the
extracted-model points  for the two  most extreme focus  levels, $f$=1
and  $f$=11 in  green and  cyan, respectively.   These green  and cyan
curves are almost everywhere lower  than the dark-blue curve, which is
consistent with them being much more  out of focus.  The central value
of the PSF in the extreme focus curves varies from 0.13 to 0.18.

\begin{figure*}
\centering
\includegraphics[width=17cm]{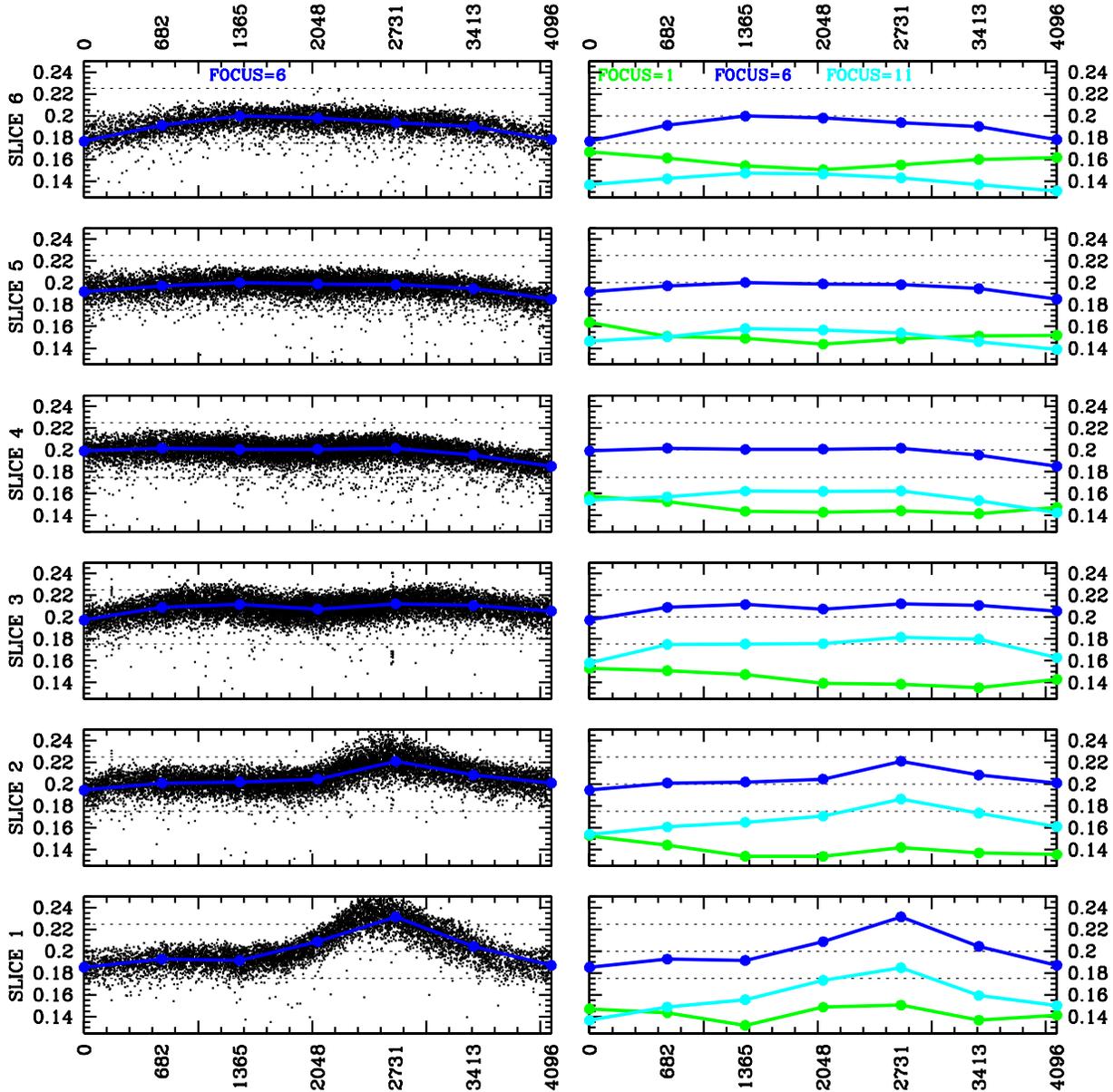}
\caption{The left plot shows the estimates of the central value of the
  PSF as a function of $x$ within 6 horizontal slices across the chips
  (as shown in the previous figure).  The individual points correspond
  to estimates of  the central value of the PSF  from individual stars
  in individual  exposures that happen  to be  centered on a  pixel to
  within  $\pm$0.25 pixel  in  $x$ and  $y$.  The  units  on the  left
  therefore refer to the fraction of a star's light that would land in
  its  central pixel  if  the star  is centered  on  that pixel.   The
  individual exposures  here are all  taken from the group  with focus
  level 6 (the  middle of the focus range).  The  solid blue dots show
  the value  of the function  for the  fiducial locations of  the PSF.
  The line  simply shows the  linearly interpolated value  between the
  fiducial  locations.   The right  plot  shows  these blue  (focus=6)
  points along with the points for  the most extreme focus location on
  one  side of  focus (green,  focus=1) and  the other  side of  focus
  (cyan, focus=11).}
\label{fig06}
\end{figure*}

\begin{figure}
\centering
\includegraphics[width=8.4cm]{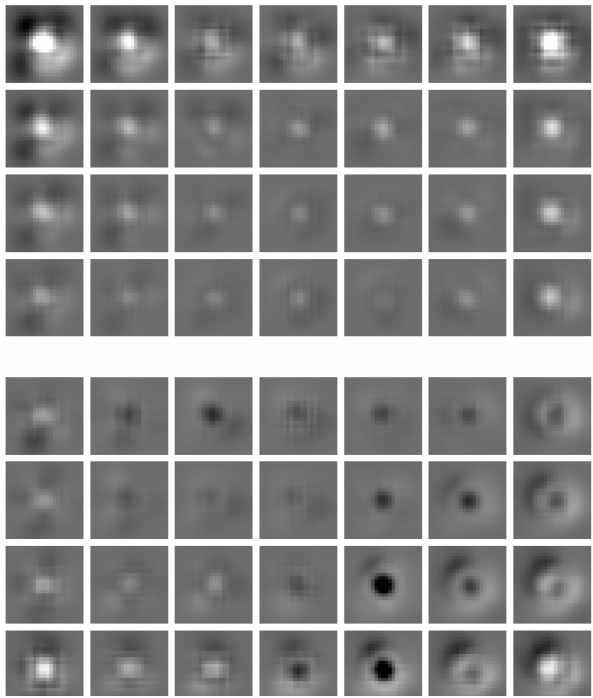}
\caption{The  array of  PSFs for  the middle  focus position  ($f$=6),
  shown with respect to the average across the detector for this focus
  level.  Black corresponds to more flux.}
\label{fig07}
\end{figure}

The previous  figure showed how  the central  pixel of the  PSF varies
with position and focus.   Figure~\ref{fig07} shows the entire central
region of the PSF for the middle  focus level ($f=6$).  In each of the
7$\times$8 panels we show  the inner 5$\times$5 flc-image-pixel region
of the PSF  (the inner 21$\times$21 PSF gridpoints) in  terms of their
residual  with respect  to the  average  PSF across  the detector  for
$f=6$.   Black  corresponds  to  more  flux  than  average  and  white
corresponds to  less flux.  It  is clear that  there is a  large sweet
spot in  the middle of  the detector, and  the PSF becomes  less tight
towards the edges of the field, particularly in the upper-left corner,
which we  know to be  very sensitive to  changes in focus.   The sharp
``happy bunny'' feature from SB13 at the bottom is also clear.

The next  figure, Figure~\ref{fig08},  shows how  the PSF  varies with
focus for 7 different locations on  the detector.  It is clear that in
the center  of the  detectors (the  top two  panels) the  middle focus
position has the  sharpest PSF, and the PSF gets  worse on either side
of focus.  This is true for  the ``happy bunny'' location as well (see
the bottom  row).  The upper  left and right  corners appear to  be in
their best focus closer to the $f=1$ extreme focus position.

\begin{figure}
\centering
\includegraphics[width=8.4cm]{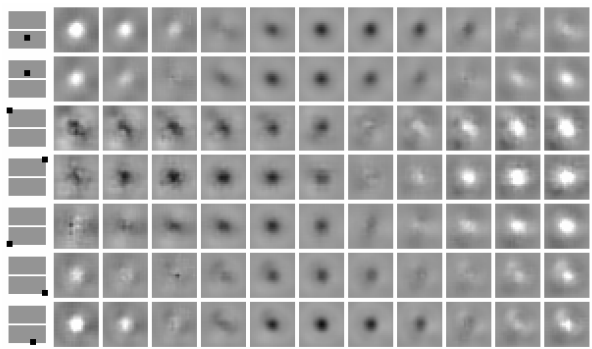}
\caption{Each  row  shows the  PSF  at  a  different location  on  the
  detector, as  indicated in  the schematic at  the left.   The panels
  from left  to right  then show the  central region of  the PSF  as a
  function of focus  level, with focus level 1 in  the leftmost column
  after the schematic and focus level 11 in the rightmost column.  The
  PSFs are shown relative to the average over all focus levels at that
  location.  Black corresponds to more flux.}
\label{fig08}
\end{figure}

It is  clear from all  of this that  the PSF behavior  is complicated,
both spatially and as a function of focus.  There is no obvious way to
reduce this  from a  3-parameter family  (x,y,f) to  anything simpler.
Even  so,  the behavior  with  focus  and  position  do appear  to  be
adequately characterized by our empirical modeling.

We have saved our focus-diverse PSF model in a simple four-dimensional
fits image that is 101 $\times$ 101 $\times$ 56 $\times$ 11, where the
first two  dimensions correspond to  the PSF model itself.   The third
dimension (56) corresponds to the 7$\times$8 array of fiducial models,
and the  fourth dimension (11)  corresponds to the focus  level.  This
allows us to construct a PSF for  any star at any location of an image
with any given focus level.


\section{Finding the PSF for each observation}
\label{sec05}
Now that we have a focus-diverse PSF model, we can fit it to the stars
in a given  exposure in order to empirically determine  the best focus
for that  exposure and  consequently the best  PSF for  that exposure.
The fact that  we have a full spatially variable  model for each focus
level means that we  can use stars from all over  the detector to help
us determine the focus.

\subsection{Using many stars to solve for the focus}
\label{ss05a}
To determine the focus for each exposure, we identified a set of about
1000 bright isolated  stars in the image.  We then  fit each star with
the model PSF that is appropriate  for its location on the detector at
a range of focus levels between 1  and 11, stepping by 0.2 focus level
(we used linear interpolation to get the PSF between focus levels).

For each fitted star at a given focus level, we determined the optimal
(x,y) position  and flux and  then got an  estimate of the  quality of
fit,  which is  simply the  absolute  value of  the residuals  between
observed  pixels and  PSF  model,  scaled by  the  flux  of the  star.
Well-fitted  stars  tend to  have  residuals  of  less than  3\%.   We
determined an  average quality  of fit  for the  stars for  each focus
level  and identified  the best  focus as  the one  that produced  the
smallest average residual for all the stars in the image.

\begin{figure}
\centering
\includegraphics[width=8.9cm]{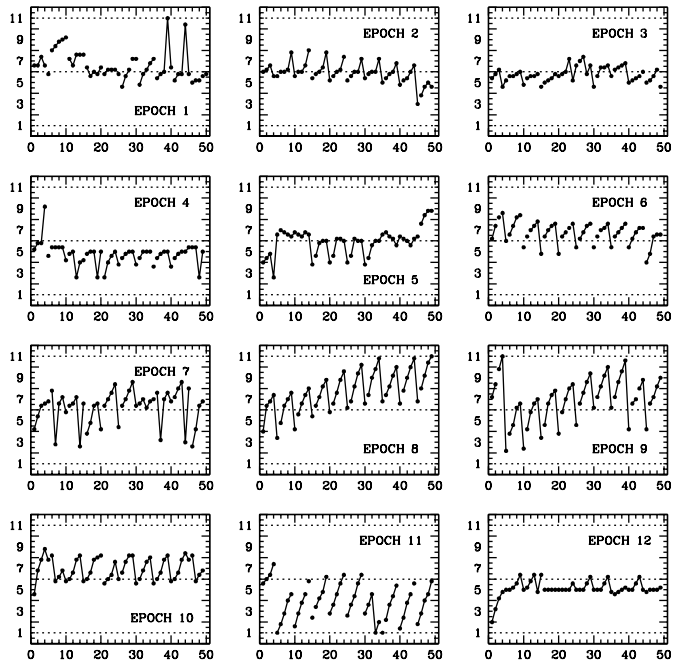}
\caption{In each  of the  twelve panels, we  show the  extracted focus
  determination   for  the   $\sim$50  exposures   taken  during   the
  corresponding epoch.  Exposures  are connected with a  solid line if
  they are taken within 500 s of each other.}
\label{fig09}
\end{figure}

We thus determined  an empirical focus level for each  exposure with a
precision of about a fifth of a focus level.  Figure~\ref{fig09} shows
the focus level  determined for the time-sorted  exposures within each
epoch in  the twelve  labeled panels.   We connect  with a  solid line
those exposures  taken within  500 seconds  of each  other (indicating
that they were taken one after  another).  We reiterate that the focus
level for  each exposure  was determined entirely  independently, thus
the trends we see reflect real and regular PSF variations.

Some of  the epochs exhibit  extremely regular variations  that repeat
every  orbit.   Others  show   less  regularity.   Clearly  the  focus
variations depend both  on the telescope's insolation  history and how
the Sun  heats it  during its  on-source pointing.   It is  beyond the
scope of this study to try  to understand these variations, rather our
aim is  simply to measure  the focus and  arrive at the  best possible
PSF.  Nevertheless, it is clear that this represents a powerful way to
examine focus changes  due to short-term breathing  or other, possibly
longer-term, phenomena such as focus drift.


\subsection{How well do individual stars predict focus?}
\label{ss05b}
Given the success of the fitting-focus  by exposure above, it is worth
asking how  many stars are needed  to pin-down the focus.   To examine
this, we identified seven images that had  focus levels of 1, 3, 5, 6,
7, 9,  and 11.  For  each exposure, we selected  500 or so  stars from
across  the  detector  that  had  S/N of  300  or  higher.   For  each
individual star,  we determined  an optimal focus  level based  on the
focus parameter  that provided the best  quality of fit to  the star's
central 5$\times$5 pixels.

\begin{figure}
\centering
\includegraphics[width=8.4cm]{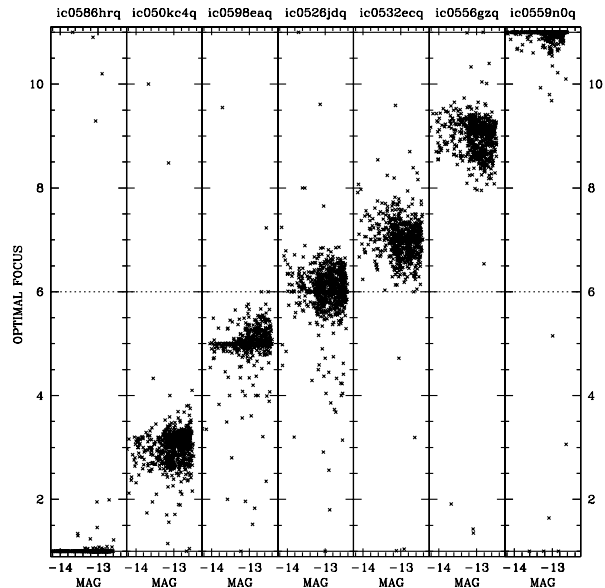}
\caption{In  each  panel,  we  show  the  focus  level  as  determined
  independently by  several hundred  extremely bright  but unsaturated
  isolated stars.   The seven panels  showcase the results  from seven
  images that span the observed focus range.}
\label{fig10}
\end{figure}

In Figure~\ref{fig10},  we show the  fitted focus  for each star  as a
function of instrumental  magnitude for the seven  chosen exposures in
seven different panels.  It is clear  that most stars in the image are
excellent predictors of the focus: typically each star can predict the
focus to within half a focus level.

\begin{figure*}
\centering
\includegraphics[width=17cm]{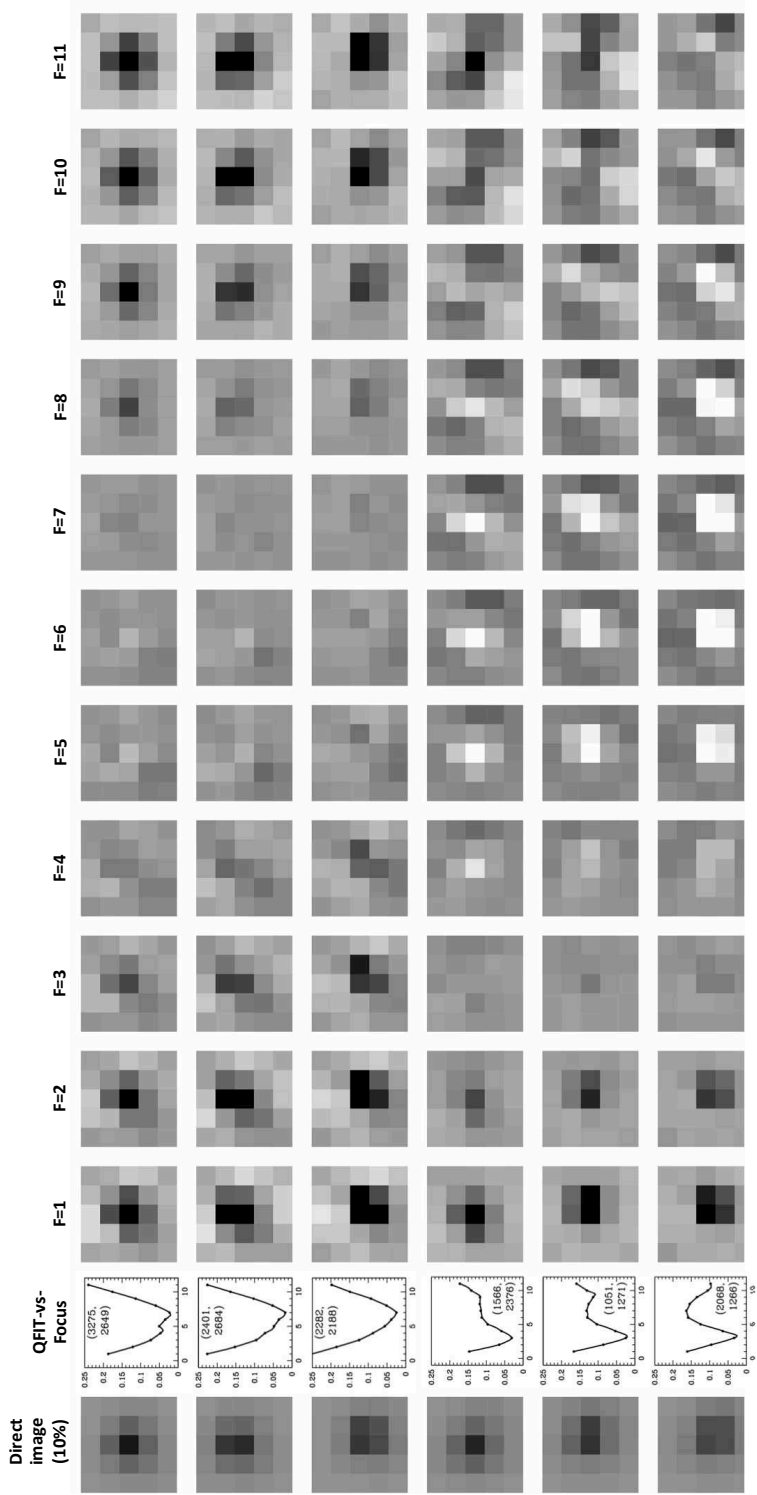}
\caption{ The  scaled fitting residuals  for six stars,  probing three
  different pixel  phases and  two images  at different  focus levels.
  The left panel shows the  5$\times$5-pixel cutout of the {\tt \_flc}
  image  centered  on  the  star.   These were  the  pixels  that  the
  focus-diverse  PSF  was  fitted  to.    The  next  panel  shows  the
  quality-of-fit  metric  as  a   function  of  assumed  focus  level.
  Finally, the eleven rasters show the scaled residuals for the eleven
  fiducial focus levels.  For each  fit, we minimized the residuals by
  allowing the center  of the star and its flux  to vary freely.  Dark
  corresponds to more flux in the image than in the model.  The direct
  star  image on  the left  is shown  scaled down  by a  factor of  10
  relative to the residuals.  The top  three rows of panels were taken
  from  exposure  {\tt  ic0532ecq}  and the  bottom  three  from  {\tt
    ic050kc4q}.
}
\label{fignew}
\end{figure*}

To  show  this  even  more   visually,  in  Fig.\ref{fignew}  we  show
explicitly the scaled residuals for six stars, probing three different
pixel phases at each of two different focus levels.  Whereas in A15 we
had to restrict our analysis to  stars that were centered on pixels in
one corner of the detector, here it is clear that with a comprehensive
model, the  variatiation with focus  can be sensed and  calibrated for
{\it all} stars at {\it all} locations on the detector.

We clearly do not need thousands of stars to pin-down the focus level.
Furthermore,  it is  not clear  from Figure~\ref{fig10}  how a  star's
signal-to-noise will affect our ability to determine focus from it.

\begin{figure}
\centering
\includegraphics[width=8.4cm]{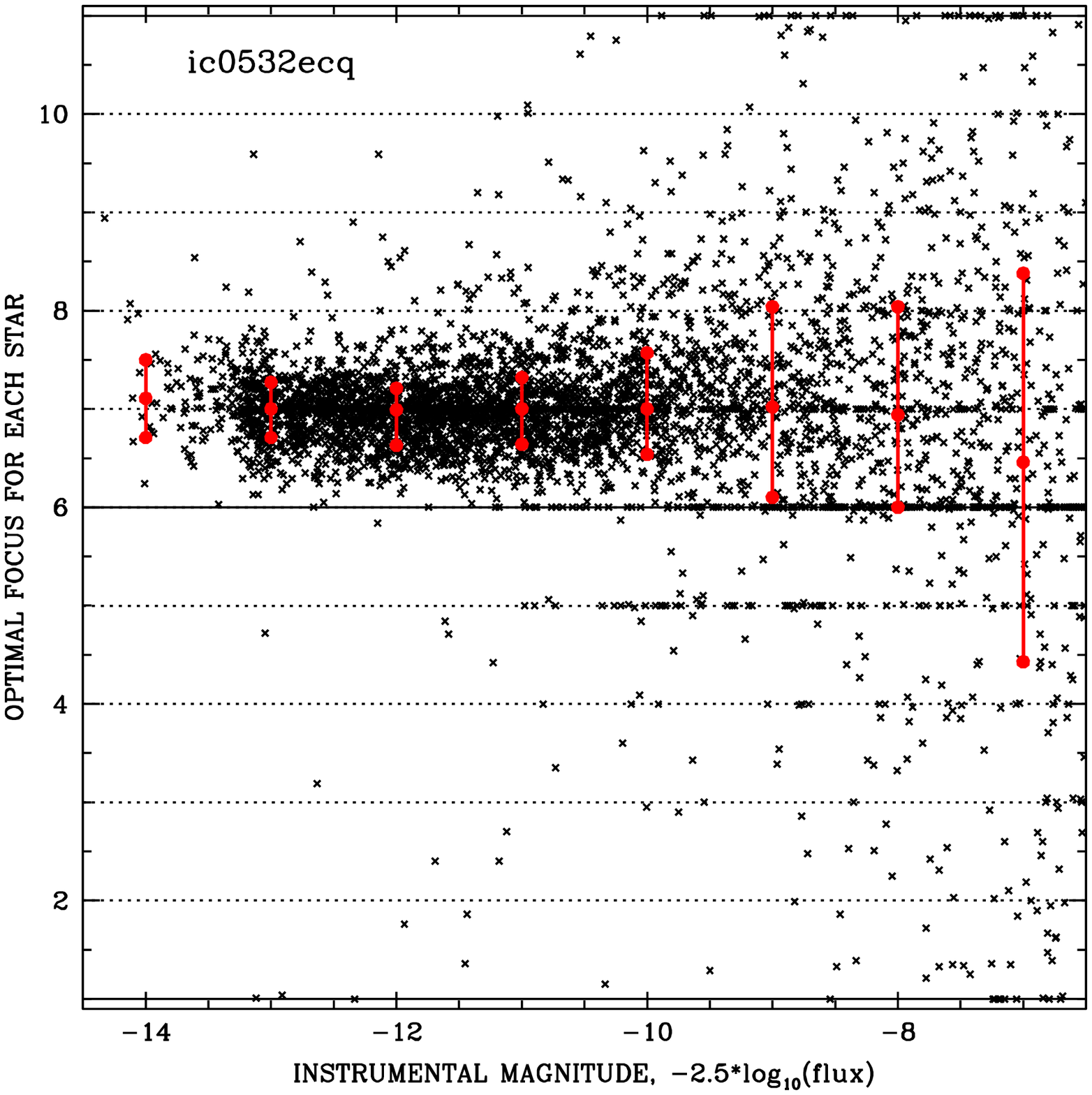}
\caption{The focus level as determined by isolated stars at a range of
  brightness  in  image  ic0532ecq  (found to  have  focus  level  7).
  Clearly the bright stars are better  predictors of focus, but even a
  few faint stars can determine the focus quite accurately.}
\label{fig11}
\end{figure}

To   explore  this,   we   took   the  image   in   the  fifth   panel
(\texttt{ic0532ecq}),  found   to  have   focus  level   $\sim$7,  and
determined the focus level for individual isolated stars with S/N from
400  ($m$=$-13$)  to  S/N  30  ($m$=$-7.5$).   The  red  errorbars  in
Figure~\ref{fig11} show  the median and $\pm$1-$\sigma$  range for the
distribution within each magnitude bin.  It is clear that a handful of
stars of moderate flux  can pin down the focus to  well within a focus
level.  This is good news, since  even sparse fields (such as the UDF)
have  about  10 stars,  several  with  reasonable S/N.   Clearly  with
focus-diverse PSF models for many more  filters it will be possible to
characterize  the focus  level for  a large  fraction of  \textit{HST}
images.

\subsection{Breathing and Platescale}
\label{ss05c}

The \textit{HST}  platescale is not constant.   As \textit{HST} orbits
the  Earth  and the  Earth  orbits  the  Sun,  its space  velocity  is
continuously changing.  Relativistic velocity aberration (VA) can lead
to changes in the platescale of up to one part in 10$^4$
(Cox \& Gilliland 2003). 
The {\tt VAFACTOR} keyword in the image headers provides a calculation
of the  VA based  on Hubble's velocity  and pointing  vectors averaged
over the exposure time.

\begin{figure}
\centering
\includegraphics[width=8.8cm]{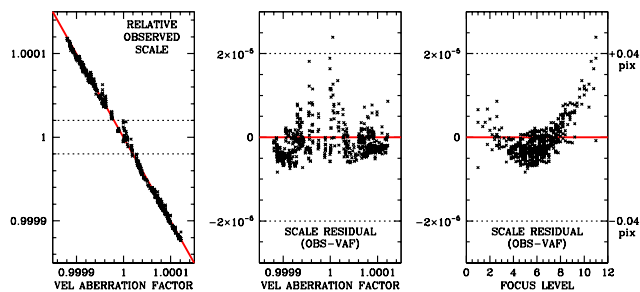}
\caption{(Left)  Observed  relative  platescale  plotted  against  the
  velocity-aberration factor keyword (\texttt{VAFACTOR}) calculated on
  the base of  the (predicted) telemetry, and provided  in the header.
  (Middle) Difference  beteen observed  and predicted  scale.  (Right)
  Difference  plotted  against  the determined  focus-level  for  each
  exposure.  The dotted lines correspond  to $\pm$0.04 pixel change at
  the edge of the detector.}
\label{fig12}
\end{figure}

In  addition to  velocity  aberration, breathing  can  also affect  an
observation's platescale, and the 589 exposures of our field give us a
unique opportunity to  examine the effect of  breathing on platescale.
For each exposure, we can  determine an average platescale by relating
the distortion-corrected positions for stars  in that frame ($x,y$) to
the positions of stars in the master frame ($X,Y$).  The platescale is
simply $\sqrt{AD-BC}$ of that global linear transformation defined by
\begin{equation}
\begin{cases}
X = A\, x + B\, y + X_\circ  \\ 
Y = C\, x + D\, y + Y_\circ. \\ 
\end{cases}
\end{equation}
The  left panel  of Figure~\ref{fig12}  plots the  observed platescale
(relative to the average) as a  function of the {\tt VAFACTOR} keyword
in  the  header  of  each  image.  Clearly,  much  of  the  platescale
variation can be explained by VA.   The middle plot shows the residual
between  the  observed  and  VA-predicted  platescale.   Finally,  the
right-most plot  shows this residual  as a function of  our determined
focus  level.  It  is clear  that the  same ``breathing''  that causes
changes  in  focus  also  causes  changes  in  the  amplitude  of  the
platescale up to 0.04-pixels.

\begin{figure}
\centering
\includegraphics[width=8.9cm]{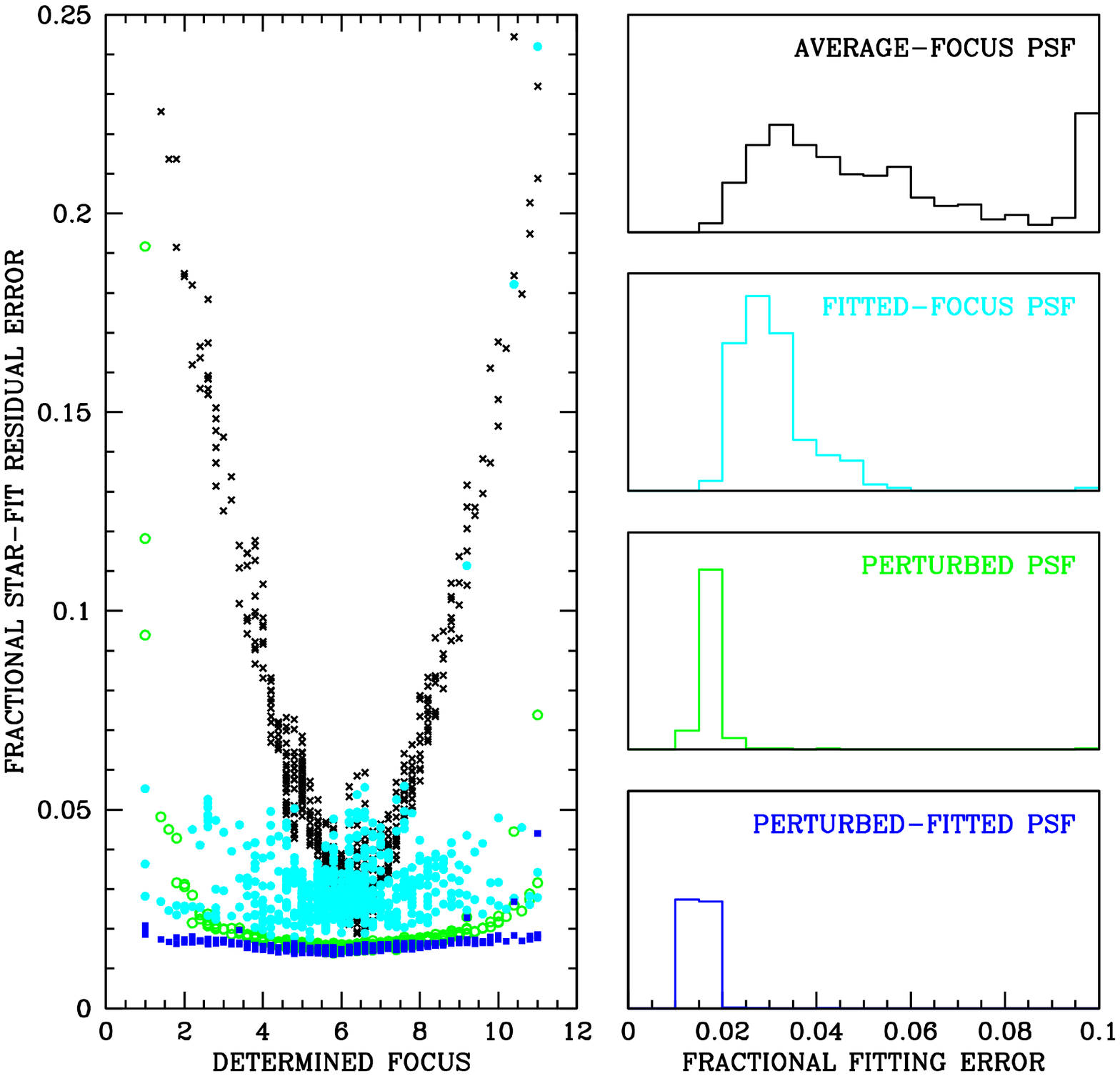}
\caption{
  (Left) The  average fractional star-fit residual  for each exposure,
  plotted  as a  function of  extracted focus  for four  different PSF
  treatments,  as labeled  on right.   (Right) Histograms  of star-fit
  residuals for the four treatments.}
\label{fig13}
\end{figure}

\section{Making use of these PSFs}
\label{sec06}

Now that we  have identified the best-focus PSF for  each exposure, it
is worth considering how well that PSF actually fits the stars.  To do
this, we investigated four different  PSFs: (1) the ``library'' PSF we
constructed for  F467M that was  designed to represent  average focus,
(2) the  fitted-focus PSF,  (3) the 4$\times$4-perturbed  library PSF,
and finally  (4) the 4$\times$4-perturbed fitted-focus  PSF.  In these
images taken at the center of M4,  we have enough stars to construct a
perturbation for  each image's  PSF.  In many  fields, we  will barely
have enough stars to pin down the  focus, meaning that there is no way
to construct a perturbation PSF for them.

There are several metrics available  to evaluate PSFs, such as quality
of fit  and photometry and  astrometry.  When we  fit a star  with the
PSF, we determine  sky from a remote annulus and  fit the star's inner
5$\times$5 pixels  with the  PSF by finding  the $(x,y)$  location and
flux that minimize the residuals in a least-squares sense, taking into
consideration the  Poisson error  in each  pixel.  Since  it is  not a
time-consuming process, we simply do a  grid search for $(x,y)$ and at
each trial position, determine the  flux by simple aperture photometry
(knowing from  the PSF what fraction  of the star's light  should fall
within the  aperture based  on its  trial position).   Once we  have a
best-fit position,  we determine  the total absolute  residual between
the  observations and  model and  divide by  the total  flux to  get a
scaled residual.

\begin{figure*}
\centering
\includegraphics[width=17cm]{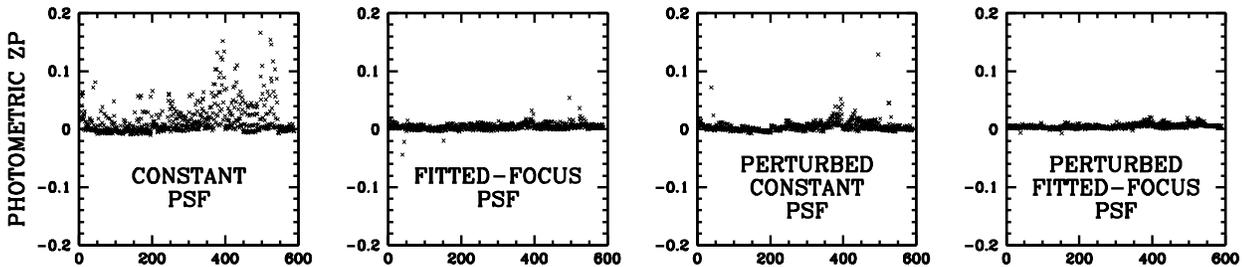}
\caption{  The empirical  photometric  zeropoint shifts  for the  four
  different PSF treatments plotted against exposure number. }
\label{fig14}
\end{figure*}

Figure~\ref{fig13} shows  the average fitting residual  for the bright
stars  within  1.5  magnitudes  of  saturation for  each  of  the  589
exposures as a function of fitted  focus for that exposure.  The black
points show the fitting error  for the unperturbed temporally constant
``library''  PSF.   The  cyan  points   show  the  residuals  for  the
fitted-focus PSF.   The green points  show the perturbed  library PSF,
and the blue  points for the perturbed focus-fitted PSF.   It is clear
that the blue  points have fitting residuals smaller than  2\% for all
focus levels.   The perturbed library  PSF is  almost as good,  but it
loses  quality  when  the  focus is  considerably  off  nominal.   The
focus-fit PSF has  residuals of about 3\% everywhere,  and the library
PSF has about  4\% errors when the  focus is good, but has  a trail to
well over  20\% errors when the  focus is off.  The  histograms on the
right show the same data summed over all focus levels.

One of the  most obvious effects of breathing is  that the fraction of
light  in the  core of  the PSF  changes.  When  fluxes for  stars are
determined by fitting a ``library'' PSF to the core, breathing results
in photometric zero-point shifts  from exposure to exposure, typically
$\pm$0.03 magnitude, but sometimes more.  Figure~\ref{fig14} shows the
measured  photometric zeropoint  shift between  each exposure  and the
average  for  the 589  F467M  exposures  for  the four  different  PSF
approaches  adopted.    The  constant  ``library''  PSF   has  typical
zeropoint shifts of about $\pm$0.02 magnitude, but they can be greater
than 0.15  mag.  The other  PSF approaches have shifts  generally less
than 0.01 magnitude,  though it is worth noting  that the fitted-focus
PSF does better than the perturbed library PSF.  Not surprisingly, the
perturbed fitted-focus PSF does best.

\begin{figure}
\centering
\includegraphics[width=8.9cm]{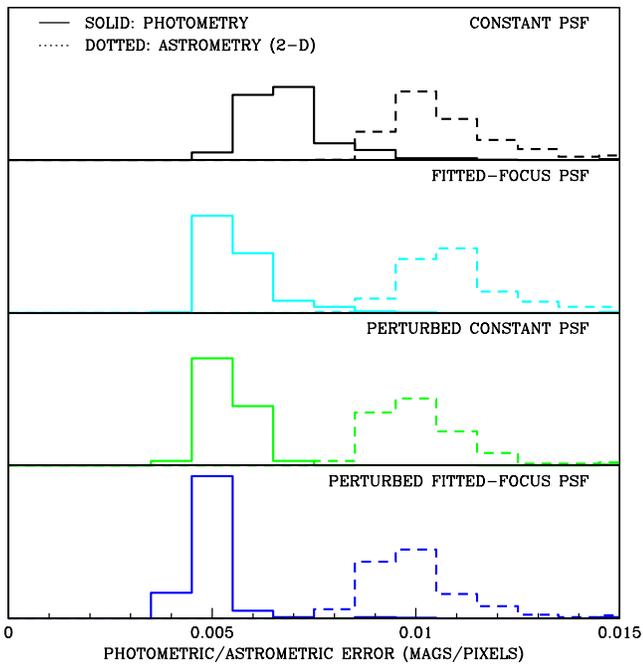}
\caption{Distribution  of RMS  photometric residuals  (solid) and  2-D
  astrometric   residuals  (dotted)   for  the   four  different   PSF
  treatments.}
\label{fig15}
\end{figure}

The photometric and astrometric residuals are the most basic test of a
PSF.  Figure\,\ref{fig15}  shows histograms  of the  photometric (solid)
and  2-D   astrometric  (dotted)  RMS  residuals   for  the  brightest
unsaturated  stars.   In  computing  these  residuals,  we  have  each
exposure to have an arbitrary photometric zeropoint and have allowed a
general    6-parameter   linear    transformation    to   match    the
distortion-corrected frame of each image to the reference frame.

It is not surprising that if  we allow for an arbitrary zeropoint, the
photometry does not  improve much with the improved  PSF. (It improves
by 15\% from 0.0065 magnitude to 0.005 magnitude RMS).  This may sound
surprising,  since  Figure 9  showed  a  20\%  variation in  the  core
intensity  with focus  changes.  Clearly  most of  the variation  with
focus involves  shifting within the 5$\times$5-pixel  aperture we used
(say, from the  core to the first diffraction  ring).  Some variation,
though, does shift light from  within the 5$\times$5-pixel aperture to
outside the aperture,  and this appears to happen  similarly for stars
across  the detector,  such that  a  single zeropoint  shift for  each
filter addresses much of it.

We  see a  similar effect  with the  astrometry.  If  we allow  for an
arbitrary scale  change (included in the  linear transformation), then
the astrometry is  not much different for the four  different PSFs: we
achieve about  0.007-pixel precision  per coordinate.   This indicates
that the shape of the PSF core does not change much with focus.

The slight improvement  of the astrometry is actually  good news.  The
original  ``library'' (average  focus) PSF  was constructed  carefully
from a  dithered set  of data  so that the  undersampled PSF  could be
properly derived free of pixel-phase bias (see Section 2.2 of Anderson
\& King 2000 for a discussion).   This PSF was used for the extraction
positions  in   our  PSF  reconstructions  here   with  no  additional
positional constraints, so it is  reassuring that the PSFs constructed
are even slightly better in an astrometric sense.  It is also the case
that the distortion  solution was derived based  on positions measured
with the ``library'' PSF, and it  is clear that the positions measured
with the focus-diverse  PSFs (and various perturbed PSFs)  are no less
accurate in a systematic sense.

\begin{figure*}
\centering
\includegraphics[width=17cm]{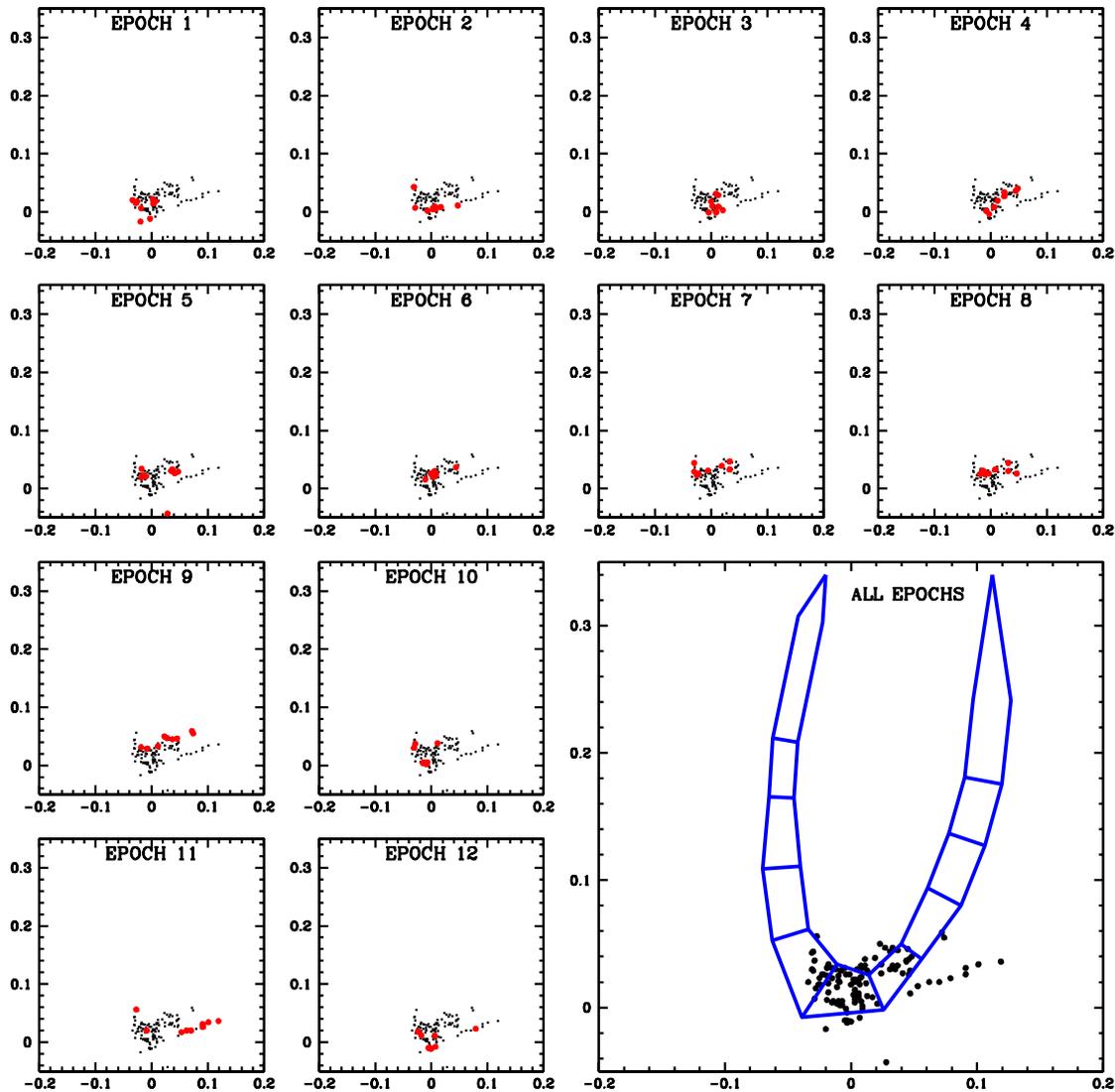}
\caption{ The ``phylogram'' plot for F775W.   It is clear that the PSF
  does not vary nearly as much for this redder filter.}
\label{fig16}
\end{figure*}

\section{F775W PSF}
\label{sec07}

Program GO-12911 also included a short  20s F775W exposure in nine out
of ten of the orbits available within each of the twelve epochs, for a
total of 108 short F775W exposures.   This is enough for a preliminary
examination of  the variation in  that PSF.   As above with  F467M, we
used a  ``library'' F775W PSF  to fit the  stars in each  exposure and
developed a 4$\times$4 array of perturbation PSFs to match the library
PSF to the image PSF.

Figure~\ref{fig16} is the F775W analog for F467M's Figure~\ref{fig03}.
The  lower-right panel  shows  the ``phylogram''  plot  for the  F775W
images, with  the regions from F467M  marked in blue.  As  before, the
distance between PSFs  corresponds to the average amount  of flux that
would have to be redistributed to  get from one to the other.  Whereas
we would have to arrange up to 30\%  of the flux to get from one F467M
PSF to another focus level, we would have to rearrange at most 10\% of
the F775W PSF.

An additional consideration  that arises when modeling the  PSFs for a
wide-band filter  is the fact  that blue star  and red stars  can have
somewhat  different  effective  wavelengths  due  to  their  different
spectral-energy distributions (SEDs).  If we  are able to come up with
a  focus-diverse  set of  PSFs  for  wide-band  filters, it  might  be
possible to introduce  an additional parameter to deal  with the color
of the stars, but that is beyond the scope of this effort.

\section{Conclusions}
\label{sec08}

We have  made use of  the uniform dataset  of GO-12911 to  explore and
model the variation  of the F467M PSF in great  detail.  We have shown
that,  by and  large,  the  PSF varies  in  predictable  ways along  a
single-parameter curve, where  the single parameter is  related to the
telescope focus.

We  have grouped  together images  that  were taken  at similar  focus
levels and  have constructed for  each of  eleven focus levels  a full
spatially variable  representation of  the PSF (i.e.,  with 7$\times$4
across each of the two 4096$\times$2048 chips).  We have examined this
PSF and find that---as expected---at  nominal focus (where most of the
exposures are taken), the PSF is sharpest at the centers of the chips.
For  the  corners of  the  chips,  the  PSF  is actually  sharpest  at
off-nominal focus positions.

Although in  a previous work  (Anderson et al  2015), we made  use the
fact that  the upper-left  corner is  particularly sensitive  to focus
changes in  order to study PSF  changes with focus, we  find here that
the  PSF across  the entire  detector is  actually quite  sensitive to
focus.  If  we fit a  typical, bright (S/N$>$50)  star with PSFs  at a
range of  trial focus levels, we  find that we can  identify the image
focus  to better  than 1  focus level  (out of  eleven).  With  only a
handful of stars, it should be possible to pin-down the focus level of
{\it  any}  exposure.  This  will  make  it  possible to  construct  a
tailor-made PSF  for every deep  exposure, since a typical  deep field
has at least five stars with reasonable signal.

The PSF model we constructed here is for F467M, admittedly an uncommon
filter.  However the procedure we developed can be used for the entire
archive: it does not depend on the fact that we are observing the same
field  in all  the  exposures.   Therefore it  should  be possible  to
construct a  focus-diverse PSF model  for all of the  common WFC3/UVIS
filters.  The WFC3/UVIS instrument team is currently pursuing this, as
such  a  detailed  understanding  of  focus  changes  will  allow  the
engineers to  make better determinations  of when to  re-adjust focus.
Having access to  a robust empirical focus measurement  will also help
engineers at the \textit{Space Telescope  Institute} to develop a more
accurate model of  how telemetry and environmental data  may be better
able to predict focus variations.

Overall, we find  that the focus-diverse PSFs  represent a significant
improvement in  the quality  of fit  for sources  and in  the absolute
photometric zeropoint  for exposures, the improvement  is particularly
significant when the telescope is out of focus.
The  new  PSFs  also  represent  a  modest  ---although  measurable---
improvement in  the photometric  and astrometric precision.   As such,
they will clearly make  the biggest difference for point-source/galaxy
discrimination, detailed PSF-fitting to  resolved or slightly resolved
objects, and for absolute-catalog work.

The particular M\,4 project at hand (GO-12911), which has enabled this
detailed  PSF study,  will benefit  of  this improved  PSFs, not  only
directly trough  the marginally  improved astrometric  and photometric
precision, but also because of the significant improvement in the PSFs
shape,  which will  help to  better disentangle  blends from  isolated
stars.   Therefore, we  will use  the perturbed  focus-diverse in  our
high-precision astrometric wobble analysis, which is the next step for
the project.

\section*{acknowledgments}
\small  JA  acknowledges  support  from  STScI  grant  GO-12911.   LRB
acknowledges  PRIN-INAF  2012  funding  under  the  project  entitled:
\textit{`The M 4 Core Project with Hubble Space Telescope'}.  We would
like to  thank Elena Sabbi,  Linda Dressel, Kailash Sahu,  and Matthew
Bourque for many useful discussions in the course of this work.

%


\label{lastpage}


\end{document}